\documentclass[12pt,a4,portrait]{article}
\usepackage{graphicx}
\usepackage{hyperref}

\newcommand{\be}{\begin{equation}}
\newcommand{\ee}{\end{equation}}
\newcommand{\beq}{\begin{eqnarray}}
\newcommand{\eeq}{\end{eqnarray}}
\newcommand{\bed}{\begin{displaymath}}
\newcommand{\eed}{\end{displaymath}}
\newcommand{\bc}{\begin{center}}
\newcommand{\ec}{\end{center}}
\newcommand{\bi}{\begin{itemize}}
\newcommand{\ei}{\end{itemize}}
\begin{document}
\baselineskip 1.2\baselineskip
\thispagestyle{empty}
\begin{center}

\vspace{2.0in}

{\Large\bf{The dynamics of scalar-tensor cosmology
\vspace{3mm}\\from RS two-brane model}}

\vspace{1.0 in}

Laur J\"arv,$^{1}$ Piret Kuusk,$^{2}$ Margus Saal$^{3}$ \\
{\it Institute of Physics, University of Tartu,
Riia 142, Tartu 51014, Estonia}\\

\end{center}

\vspace{0.5in}

\begin{abstract}
We consider a Randall-Sundrum two-brane cosmological model
in the low energy gradient expansion approximation by Kanno and Soda. 
It is a scalar-tensor theory with a specific coupling function 
and a specific potential. Upon introducing the FLRW metric
and perfect fluid matter on both branes in the Jordan frame, 
the effective dynamical equation for the the A-brane (our Universe) 
scale factor decouples from the scalar field and B-brane matter
leaving only a non-vanishing integration constant
(the dark radiation term). 
We find exact solutions for the A-brane scale factor for four types 
of matter: cosmological constant, radiation, dust, and cosmological 
constant plus radiation. 
We perform a complementary analysis of the dynamics of the scalar
field (radion) using phase space methods and examine convergence 
towards the limit of general relativity. 
In particular, we find that radion stabilizes at a certain  finite value 
for suitable negative matter densities on the B-brane.
Observational constraints from Solar system experiments (PPN) and
primordial nucleosynthesis (BBN) are also briefly discussed. 
\bigskip


\end{abstract}

\vspace{3in}
$^1$ Electronic address:  laur@fi.tartu.ee

$^2$ Electronic address: piret@fi.tartu.ee

$^3$ Electronic address: margus@hexagon.fi.tartu.ee


\newpage

\section{Introduction}


A great challenge for contemporary theoretical cosmology
is to find and examine possible frameworks for explaining new observational
data (dark energy, accelerated expansion, etc.) 
which may not be accommodated in the standard cosmological
model based on the Friedmann solution of the Einstein equations.
Investigations of alternative cosmological models include
the braneworld scenario, where our visible
Universe is one of the two hypersurfaces embedded 
in a 5-dimensional bulk spacetime, while
the proper distance between the branes  
is measured by the radion field.

The first phenomenological braneworld models 
were presented by Randall and Sundrum \cite{rs1, rs2} in order to 
solve the hierarchy problem of the 
Standard Model of particle physics. A more fundamental  
theoretical setup for braneworlds arises
in the 10-dimensional superstring theory 
(or the 11-dimensional M-theory) with  5 
(6) compactified dimensions leaving us a 5-dimensional 
cosmological bulk spacetime \cite{lukas}. 
In models with several branes which move in the bulk, 
a collision of the branes can 
mimic the big bang \cite{ekpyrotic} or the exit from inflation \cite{ksw}.
There are at least two approaches to get a 4-dimensional 
effective theory from the 5-dimensional theory.
The first one, developed by Kanno and Soda \cite{ks2, ks1}, is the gradient 
expansion method 
where the brane tension is taken to be much bigger 
than the energy density on the 
brane. 
The second approach, 
developed by Brax et al. \cite{bbdr}, is known as the moduli space 
approximation. In this approach the brane 
positions are described by scalar moduli fields 
emerging from compactification assuming 
that the motion of the branes is slow. 
It can be shown that at least at the first order both approximations
agree \cite{kanno}.

Following Kanno and Soda \cite{ks1}, we consider 
the Randall-Sundrum type I cosmological scenario \cite{rs1} 
with two branes (A and B) moving in a 5-dimensional bulk.
Both branes are taken to be homogeneous and isotropic, 
and supporting energy-momentum tensors of
a perfect barotropic fluid, with barotropic index $\Gamma$
on the A-brane (identified with our visible Universe) 
and $\Gamma^{^B}$ on the B-brane. 
The field equations of the effective 4-dimensional theory
 obtained by Kanno and Soda \cite{ks1} are
the equations of a scalar-tensor theory
with one scalar field (interpreted as a radion) 
with a specific coupling function  $\omega(\Psi)$.

In recent decades, a lot of work has been done on 
studying scalar-tensor cosmologies with non-constant
coupling function $\omega(\Psi)$. The general properties of 
the evolution of $\Psi$ in the 
Einstein frame were investigated by Damour and Nordtvedt \cite{dn} and 
in the Jordan frame by Serna et al. \cite{sa1},
while an analogous study in the case of a massive scalar
field was presented by Santiago et al. \cite{skw2}.
A comprehensive analysis of the early and late time cosmological
models in the Jordan frame was carried out by Barrow and Parsons \cite{bp}.
Various observational constraints on scalar-tensor cosmologies 
were discussed by Santiago et al. \cite{skw1}, 
Damour and Pichon \cite{dp}, Clifton et al. \cite{cbs},
and Coc et al. \cite{coc}, and for models with two branes
by Palma \cite{palma}.
In these papers, most calculations are performed
in the unphysical Einstein frame, although the Jordan frame
is used for physical interpretations.\footnote{For a lucid explanation
of the relevance and relationship of the two frames, as well as a 
general review of scalar-tensor cosmology, see the 
recent book by Faraoni \cite{Faraoni_book}.}

The scalar-tensor
cosmological model we consider in the Jordan frame
stands out from those mentioned above in three respects. First, 
the second brane can contain matter and its matter tensor 
acts as an additional source  in the equations of the gravitational 
and the scalar field. Second, the coupling function of
the scalar field has a specific form 
$\omega (\Psi) = \frac{3 \Psi}{2 (1 - \Psi)}$ which is fixed
by the construction of the model. Third, there are possible
cosmological constants on both branes 
appearing via a specific potential of the scalar field, so 
in general we consider two-component source terms (matter + cosmological
constants on both branes).

The cosmology of this model was first 
investigated by Kanno et al. \cite{kss} and 
later also by us \cite{iii}.\footnote{The cosmology and radion dynamics 
of RS two-brane model has been studied my many authors 
using various approaches \cite{RS_cosmo_all}, the closest 
to ours being the moduli space approximation 
\cite{bbdr, palma_davis_MSA, webster_davis_MSA}.}
In this paper we obtain several new general analytic solutions
for the flat ($k=0$) A-brane scale factor, thus encompassing solutions
for all interesting matter tensors: 
1) cosmological constants and no matter, 
2) radiative matter and no cosmological constants, 
3) dust matter and no  cosmological constants, as well as
4) radiative matter and cosmological constants.
We also find the accompanying analytic solutions for the radion field
in cases 2) and 4) above.
We apply the methods of dynamical systems 
to study the evolution of the radion field,
finding fixed points and 
drawing phase portraits related to the first three cases. 
The global picture emerging from this analysis is more clear
and has some advantages over the usual procedure 
of studying the convergence of scalar-tensor theories
towards general relativity,
which relies on an analogy with a fictitious particle 
with velocity dependent effective mass  \cite{dn, sa1, skw2}.
Depending on the type and the (initial) density of A- 
and B-brane matter,
the branes typically evolve towards infinite separation
or to a collision, but for certain negative B-brane
energy densities, the radion (corresponding to the inter-brane distance)
is stabilized by an attractor at a certain finite value.
Finally, we consider Solar system experiments and
Big Bang nucleosynthesis to derive some observational limits
on the parameters of the model.

The plan of the paper is as follows. Sects. 2 and 3 
give the basic equations of the model
and clarify their general properties. 
Sect. 4 presents the analytic solutions, while 
Sect. 5 carries out a complementary 
analysis of the dynamics of the scalar field. 
Sect. 6 follows with a brief discussion on observational constraints, 
and Sect. 7 contains some concluding remarks.


\section{Effective action and field equations in the Jordan frame}


The starting point is the 5-dimensional Randall-Sundrum action \cite{rs1}
\beq 
I_{_{RS}} = \frac{1}{2\kappa_{5}^2} \int d^5 x \sqrt{-g} \left( {\cal{R}} - 2 
\Lambda_{5} \right) 
\label{rs1a}
+ \int d^4 x \sqrt{-h} \left ( {\cal L}_{m} - \Sigma \right) 
+ \int d^4 x \sqrt{-f} \left ( {\cal L}^{^B}_{m} - \Sigma^{^B} \right) \, ,
\eeq
where $\cal{R}$ is the scalar curvature, while $h_{\mu\nu}$ and 
$f_{\mu\nu}$ 
are the induced metrics on the A- and B-brane, respectively. 
We assume that the brane tensions can be divided into two parts
\be 
\Sigma = \sigma_{0} + \sigma \ , \qquad \quad 
\Sigma^{^B} = \sigma_{0}^{_B} + \sigma^{_B} \ , 
\ee
with $|\sigma_{0} | \gg |\sigma|$, 
$|\sigma_{0}^{_B} | \gg |\sigma^{_B}|$.
The first part is balanced with the bulk cosmological constant $\Lambda_{5}$,
\be
(\sigma_{0})^2 = (\sigma^{_B}_{0})^2 = -\frac{6 
\Lambda_{5}}{\kappa^{2}_{5}}\ , \qquad \qquad 
\Lambda_{5} \equiv -\frac{6}{\ell^2}\, , 
\ee
where $\ell$ is the curvature radius of 5-dimensional bulk anti-de Sitter 
spacetime.
The second part gives rise to  unbalanced energy density on the 
branes which can be interpreted
in terms of effective cosmological constants
  $\lambda$, $\lambda^{^{B}}$ as
\be
\sigma = \frac{\lambda\,c^{2}}{\kappa^{2}}\, , \qquad
\sigma^{_{B}} = \frac{\lambda^{^{B}}\,c^{2}}{\kappa^{2}} \, .
\ee

Let us introduce a dimensionless scalar field $\Psi$, called radion, 
which measures the proper orthogonal distance 
between the branes situated at $y = 0$ and $y = \ell$ (at the fixed
points of $S_1 / Z_2$ orbifold spacetime), 
\be \label{pd}
d(x) \equiv \int_{0}^{\ell} \sqrt{g_{_{55}}}\ dy 
= -\frac{\ell}{2} \ln (1 - \Psi)\,,  \qquad \Psi \in [0,1]\,,
\ee
in coordinates where $g_{55}= g_{55} \ (x)$.
Then, according to the low energy expansion scheme developed 
by Kanno and Soda, 
in a sense that the energy density of the matter on the brane is smaller than 
the brane tension $|\rho/\Sigma| \ll 1 $, 
$|\rho^{_{B}}/\Sigma^{^{B}}| \ll 1$, 
the 4-dimensional field equations on the A-brane can 
be written as \cite{ks1}
\beq \label{ksG}
G^\mu_{\ \nu} (h)&=& \frac{\kappa^2}{\Psi } \left[T^{A\mu}_{\quad \nu}(h) + (1-
\Psi )^{2} \, T^{B\mu}_{\quad \nu}(f) \right]
-\frac{\kappa^2}{\Psi} \delta^{\mu}_{\nu} \left[\sigma + 
(1-\Psi)^{2} \, \sigma^{_B} \right] \nonumber \\
&&+{ 1 \over \Psi } \left(  \Psi^{|\mu}_{\ |\nu}  -\delta^\mu_\nu \, 
\Psi^{|\alpha}_{\ |\alpha} \right) 
+{\omega(\Psi ) \over \Psi^2} \left( \Psi^{|\mu}  \Psi_{|\nu}
-{1\over 2} \delta^\mu_\nu \, \Psi^{|\alpha} \Psi_{|\alpha} \right)  \ ,
\\
\label{ksP}
\Psi^{|\mu}_{\ |\mu} &=& \frac{\kappa^2}{2\omega + 3} 
\left[ T^{^A}(h) + (1-\Psi)\, T^{^B}(f) \right]
-  \frac{4 \kappa^2}{2\omega + 3}  \left[ \sigma + 
(1-\Psi) \, \sigma^{_B} \right] \nonumber\\
&&-\frac{1}{2\omega +3} \frac{d\omega}{d\Psi} \Psi^{|\mu} \Psi_{|\mu}\,,
\eeq 
where $^|$ denotes the covariant derivative with respect to the A-brane metric 
$h_{\mu\nu}$ and $\kappa^2 \equiv \kappa^2_5 /\ell$ is the 4-dimensional
bare gravitational constant. 
The brane metrics are conformally related by $f_{\mu\nu}=(1-\Psi) h_{\mu\nu}$
while the function
\be \label{ksw}
\omega(\Psi) \equiv \frac{3}{2} \frac{\Psi}{1 - \Psi}   \,
\ee
describes the proper distance $d$ between the branes as
(cf. \cite{garriga})
\be
\omega = 
\frac{3}{2} \ \left( e^{2 \frac{d}{\ell}} - 1 \right) \,. 
\ee

It is noteworthy that Eqs. (\ref{ksG}) and (\ref{ksP}) coincide 
with the equations of
the general scalar-tensor theory with a specific form (\ref{ksw}) of the 
coupling function $\omega(\Psi)$ and an extra matter term from the B-brane. 
These equations can be 
derived from the 4-dimensional scalar-tensor action 
\beq \label{bdf4da}
I^{^A}_{_{JF}}  &=& {1 \over 2 \kappa^2} \int d^4 x \sqrt{-h}
      	        \left[ \Psi R(h) - {\omega (\Psi ) \over \Psi}
      		\Psi^{|\alpha} \Psi_{|\alpha} -2 \kappa^{2} \ V(\Psi)\right] 
\nonumber \\ \nonumber\\
        &&+ \int d^4 x \sqrt{-h} {\cal L}^{^A}_{m}
       	+ \int d^4 x \sqrt{-h} \left(1-\Psi \right)^2 {\cal L}^{^B}_{m}  \ , 
\label{4Daction}
\eeq 
where $\omega(\Psi)$ is given by (\ref{ksw}) 
and $V(\Psi)$ denotes an effective potential introduced by cosmological
constants on the branes, 
\be \label{VPsi}
V(\Psi) = {\sigma}  + {\sigma^{_B}}  (1-\Psi)^2 
\label{4Dpotential}\,.
\ee
If ${\cal{L}}^{^B}_m \equiv 0$ and $\sigma^{_B} =0$, 
the influence of the B-brane disappears from the action 
and we are left with scalar-tensor a theory with a constant potential
$V = \sigma$.
The Eqs. (\ref{ksG}) and (\ref{ksP}) reduce to Einstein's general relativity 
when ${\dot \Psi} =0$,  $\Psi = 1$, 
i.e., at $\omega(\Psi) \rightarrow \infty$,
  $d \rightarrow \infty$.


\section{FLRW type two-brane cosmology}


\subsection{Field equations}

Assuming that in the Jordan frame the Universe on the A-brane is described by 
the FLRW line element 
\beq \label{FRW}
ds^2 &=& - dt^2 + a^{2}(t) \left[\frac{dr^2}{1- k r^2} +
      r^{2} \left( d\theta^{2} + \sin^2 \theta d\varphi^{2} \right)\right] \,,
\qquad k = 0, \pm 1 \,,
\eeq
and the matter on the branes is modeled by 
a perfect barotropic fluid with respect to the corresponding brane metric
\be \label{sm:if}
T^{^A}_{\mu\nu} = (\rho + p) u_{\mu} u _{\nu} + p \, h_{\mu\nu} \,, \qquad p = 
(\Gamma -1) \rho \, , 
\ee 
\be
T^{^B}_{\mu\nu} = (\rho^{_B} + p^{_B}) u^{_B}_{\mu} u^{_B}_{\nu} + p^{_B} 
f_{\mu\nu} \,, \qquad p^{_B} = 
(\Gamma^{^B} -1) \rho^{_B} \, ,
\ee
we can write the 4-dimensional  field equations (\ref{ksG}) and (\ref{ksP}) as 
follows: 
\beq 
\label{00}
H^2 &=& 
- H \frac{\dot \Psi}{\Psi} 
+ \frac{1}{4} \frac{\dot \Psi^2}{\Psi (1 - \Psi)} + \frac{\kappa^2}{3} 
\frac{V}{\Psi} 
+ \frac{\kappa^2}{3} \frac{\rho}{\Psi} + \frac{\kappa^2}{3} \frac{(1- 
\Psi)^2}{\Psi} \rho^{_B} 
- \frac{k }{a^2} \ ,\\ \nonumber \\ 
\label{mn}
2 \dot{H} + 3H^2 &=& 
-\frac{\kappa^2}{\Psi} p - \frac{\kappa^2}{\Psi} (1-\Psi)^2 p^{_B}  + 
\frac{\kappa^2}{\Psi} V 
- \frac{\ddot{\Psi}}{\Psi} - 2 H \frac{\dot{\Psi}}{\Psi} - \frac{3}{4} 
\frac{\dot{\Psi}^2}{\Psi(1-\Psi)} 
- \frac{k }{a^2} \ , \\ \nonumber \\ 
\label{deq}
\ddot \Psi &=& - 3 H \dot \Psi - \frac{1}{2} \frac{\dot \Psi^2}{(1-\Psi)} 
+ \frac{2}{3} \kappa^2 \ \left(2 V  - \Psi \frac{dV}{d\Psi}\right)
\left( 1 - \Psi \right)
  \nonumber \\ 
&&+ \frac{\kappa^2}{3}(1-\Psi) \ ( \rho - 3p) 
+ \frac{ \kappa^2}{3}(1-\Psi)^2   ( \rho^{_B} - 3 p^{_B} )  \ . 
\eeq
In the case ${\dot \Psi} =0$, $\Psi = 1$ these reduce to  
the general relativistic FLRW equations 
with cosmological constant and barotropic perfect fluid. Note
that since cosmological constant is included separately, its
modeling by barotropic index $\Gamma =0$ is redundant.

The Bianchi identity yields the following equations (conservation laws) for 
the matter fluids:
\be \label{ceA}
\dot{\rho} + 3 H \Gamma \rho = 0
\quad \Rightarrow \quad \rho = \rho_{0} \, \left(\frac{a}{a_{0}}\right)^{-3 
\Gamma} \,,
\ee
\be  \label{ceB}
\dot{\rho}^{_B} + 3 H \Gamma^{^B} \rho^{_B} - \frac{3}{2} 
\frac{\dot{\Psi}}{(1-\Psi)} \Gamma^{^B} \rho^{_B} = 0 
\quad \Rightarrow \quad
\rho^{_B} = \rho^{_B}_{0} \left( \sqrt{\frac{1-\Psi}{1-\Psi_{0}}} \, 
\frac{a}{a_{0}} \right)^{-3 \Gamma^{^B}} \, 
\ee
(here we assume that $\rho/ \rho_0 > 0$, $\rho^{_B}/\rho^{_B}_0 >0$, 
allowing the possibility of exotic negative energy density
on the B-brane). 
Combining Eq. (\ref{ceA}) with Eq. (\ref{ceB}) we get a relation 
between the energy densities on both branes
\be \label{rhoAB}
\rho^{_B} = \rho^{_B}_{0} 
\left({\frac{1-\Psi}{1-\Psi_{0}}} \right)^{-\frac{3}{2} \Gamma^{^B}} \, 
\left( \frac{\rho}{\rho_{0}} \right)^{\frac{\ \Gamma^{^B}}{\Gamma}}\,.
\ee
The meaning of integration constants is given by Eqs. (\ref{ceA}) and 
(\ref{ceB}): at a fixed moment $t_0$ we have
$a(t_0) = a_0$, $\rho(t_0) = \rho_0$,  $\rho^{_B}(t_0) = \rho_0^{_B}$,
$\Psi(t_0) = \Psi_0$.

Combining Eqs. (\ref{00}) and (\ref{mn}) and using the radion equation 
(\ref{deq}) 
we get the dynamical equation for $H$, 
\beq \label{Heq}
\dot H + 2 H^2  =  \frac{2}{3} \kappa^2 \, \sigma + 
\frac{\kappa^2}{6}(\rho - 3 p) - \frac{k }{a^2} \,,
\eeq
which does not contain any additional terms 
compared with the usual FLRW cosmology. 
This happens due to the specific form of the coupling function 
(\ref{ksw}) and the potential (\ref{VPsi}) 
which cancel all terms 
containing radion, B-brane matter and B-brane cosmological
constant. 
Yet, the differences from the general relativistic FLRW model are
concealed in the bare gravitational constant $\kappa^2$ which
does not coincide with the Newtonian gravitational constant (see 
Sect. 6), and also in the form of the  constraint equation
(\ref{00}).

In the case $\rho \not= 3p$ ($\Gamma \not= \frac{4}{3}$)
the first integral of Eq. (\ref{Heq}) reads 
\be \label{Heq_fi}
\frac{d (a^2)}{\sqrt{\sigma a^4 + \rho_0 \ a_0^{3\Gamma} \ a^{4-3\Gamma} 
- \frac{3}{\kappa^2}\, k  \, a^2 +
  C a_0^4}} 
= 2 \sqrt{\frac{\kappa^2}{3}} dt \ ,
\ee
where $C$ is a constant of integration.
Using the conservation law (\ref{ceA}) it can be written 
as an expression for the Hubble parameter, 
\be \label{H^2}
H^2 = \frac{\kappa^2}{3} \, \sigma + \frac{\kappa^2}{3} \rho_{0} 
\left(\frac{a}{a_{0}}\right)^{-3 \Gamma} 
- \frac{k }{a^2} + 
\frac{\kappa^2 \,C}{3} \left(\frac{a}{a_0}\right)^{-4} \ .
\ee
The last term is forbidden in the case of the standard FLRW cosmology due 
to the Friedmann constraint but is allowed in the present case.
Let us note that Eq. (\ref{H^2}) coincides with the Friedmann equation 
of the fine-tuned one-brane cosmological model \cite{langl} 
in the low energy approximation where the term $\sim \rho^2$ is neglected.
The integration constant $C$ is geometrical in nature and 
can be interpreted as the value of 
the energy density of the dark radiation on the A-brane at the moment $t_0$.

In the case $\rho = 3p$ ($\Gamma = \frac{4}{3}$)
an equation corresponding to Eq. (\ref{H^2}) reads 
\be \label{H^2rad}
H^2 = \frac{\kappa^2}{3} \, \sigma  - \frac{k }{a^2} + 
\frac{\kappa^2 \,K}{3} \left(\frac{a}{a_0}\right)^{-4} \ ,
\ee
where $K$ is an integration constant which can be 
redefined as a sum of the initial value of radiative matter 
density $\rho_0$ and the initial value of the dark radiation
density $C$,  $K \equiv \rho_0 + C$. 
For $k = 0$, $\sigma = 0$ the exact analytic solutions 
for $a(t)$ and $\Psi(t)$ are presented in Subsect. 4.2.

Eqs. (\ref{Heq}), (\ref{H^2}) imply an expression
for acceleration,
\be
\frac{{\ddot a}}{a} = \frac{\kappa^2}{6} \left( (2 - 3 \Gamma) \rho
+ 2 \sigma \right) - \frac{\kappa^2 a_0^4 \ C}{3 a^4} \, ,
\ee 
which differs from the corresponding expression of the FLRW cosmology
only due to the last term. To get an additional drive for acceleration 
besides the usual cosmological constant,
we see the integration constant $C$ 
must be negative, $ C = - |C|$.
\footnote{The same phenomenon was noticed before in the studies of a single 
brane \cite{dark_acceleration}.}
However, in an expanding Universe the last term diminishes quickly in time, 
so it could have been influential in the early Universe, but is negligible 
at present.


\subsection{Two-brane cosmology in the conformal time}

In order to clarify some general properties of the model
let us introduce the conformal time on the A-brane,  
$dt = a(t) d \eta$ as usual, and let us redefine the scalar field as 
$1 - \Psi = \chi^2$. 
Then from Eqs. (\ref{00})--(\ref{deq}) the following 
equations follow:
\beq 
\label{S}
(a^\prime)^2 - ((a \chi)^\prime)^2 + k (a^2 - (a \chi)^2)  &=& 
\frac{\kappa^2}{3} a^4
(\sigma + \sigma^{_B} \chi^4 + \rho + \rho^{_B} \chi^4)
  \ ,\\ \nonumber \\ 
\label{A}
a^{\prime \prime} + ka  &=&   \frac{\kappa^2}{6} a^3 
(4 \sigma +(4 - 3 \Gamma) \rho) \ , \\ \nonumber \\ 
\label{B}
(a \chi)^{\prime \prime} + k (a\chi) &=&
-\frac{\kappa^2}{6} (a \chi)^3 (4\sigma^{_B} + (4 - 3 \Gamma^{^B})
  \rho^{_B})   \ , 
\eeq
with a prime here denoting  $d/ d\eta$. The result is not surprising
taking into account that the B-brane metric is related to the A-brane metric 
by a conformal transformation with the conformal factor $\chi^2$.

Dynamical equations (\ref{A}) and (\ref{B}) can be integrated
and their first integrals read 
\be
\label{A1}
(a^\prime)^2 =  \frac{\kappa^2}{3} a^4 
(\sigma + \rho) - k a^2 + \frac{\kappa^2\, a_0^4}{3}\, C \,,
\ee
\be
\label{B1}
((a \chi)^\prime)^2 = - \frac{\kappa^2}{3} (a \chi)^4 
(\sigma^{_B} + \rho^{_B}) - k (a \chi)^2 + C^{^B} \ ,
\ee
where  $C^{^B}$ is another constant of integration. Eq. (\ref{S})
constrains it to be 
$C^{^B} =  \frac{\kappa^2\, a_0^4}{3}\, C  $.
We see that the influence of the B-brane on the A-brane manifests itself
only in permitting non-vanishing values of $C$ which must be proportional to 
the value of the integration constant $C^{^B}$ in the solution for scale 
factor $(a\chi)$ of the B-brane. 
If $C=0$ the cosmologies on the both branes decouple and
amount to the usual FLRW models.
A non-vanishing $C$ can be interpreted as an initial value of the
energy density of the dark radiation on the A-brane (see Subsect. 2.2)


\section{Analytic solutions for a flat ($k=0$) Universe}


In general, Eqs. (\ref{00})--(\ref{deq})
describe a cosmological model with a two-component source 
term containing  the cosmological constant and 
a perfect fluid matter ($\Gamma \not= 0$) on both branes. 
Cosmologically interesting matter tensors are 
radiation ($\Gamma = 4/3$) and dust ($\Gamma = 1$). 
However, the full analytic solutions of Eqs. (\ref{00})--(\ref{deq}) 
can be found only in some special cases.


\subsection{Cosmological constant dominated Universe}

In the  case of vanishing 
matter density ($\rho = 0, p = 0 $) and positive cosmological constant
on the A-brane ($\sigma > 0$) 
the solution of Eq. (\ref{Heq}) for a flat Universe ($k=0$) 
reads\footnote{The solution with ${\rm th}$ was previously written down 
in the moduli space approximation \cite{webster_davis_MSA}.}
\be
  H  = \left\{ \begin{array}{ll}
             \ \pm \sqrt{\frac{\kappa^2 \sigma} {3}} \ 
{\rm th} \left(\pm 2\sqrt{ \frac{\kappa^2  \sigma }{3} } \ (t - t_1) \right) \\
             \ \pm \sqrt{\frac{\kappa^2 \sigma }{3}  }\ {\rm cth} 
\left(\pm 2\sqrt{ \frac{\kappa^2  \sigma}{3} } \ (t - t_1) \right)   \
                 \end{array}
        \right. 
, \quad \Longrightarrow \quad
\label{H_G4/3_k0}
  a^2  = \left\{ \begin{array}{ll}
             \ a_1^2\  {\rm ch} 
\left(\pm 2\sqrt{\frac{\kappa^2 \sigma}{3} } \
                            (t - t_1) \right)    \\
             \ a_1^2 \ \left| {\rm sh} \left(
  \pm 2 \sqrt{\frac{\kappa^2  \sigma}{3}  } \
  (t - t_1) \right) \right|
                 \end{array}
        \right. 
\ee
where $a_1$, $t_1$ are integration constants.
The solutions (\ref{H_G4/3_k0}) approach  asymptotically 
in  time the de Sitter  solution $H^2 = \frac{\kappa^2}{3}  \sigma $,
$\sigma > 0$.
In an expanding Universe ($H>0$) 
the solution  $H \sim {\rm th} (t)$ is accelerating (${\dot H} > 0$), 
while $H \sim {\rm cth} (t)$ is decelerating (${\dot H} < 0$).
The influence of the dark radiation can be seen from 
the first integral (\ref{H^2}) where integration constants
are included  differently, 
\be \label{fi} 
H^{2} = \frac{\kappa^2}{3} \left( \sigma + 
\frac{C \ a_0^4}{a^4} \right) \,.
\ee
Here for the expanding Universe the last term on RHS is quickly 
decaying away
leaving simply de Sitter expansion independent of dark radiation.\footnote{In 
this case ($C = 0$) an analytic solution also
for the scalar field was presented in ref. \cite{iii}.}

For negative cosmological constant, $\sigma<0$, the corresponding
solution reads
\be
  H  = \mp \sqrt{\frac{\kappa^2  |\sigma| }{3} }\ 
{\rm tan} \left(\pm 2\sqrt{ \frac{\kappa^2  |\sigma| }{3}} \ (t - t_1) \right) \,,
\quad \Longrightarrow \quad
  a^2  =       \ a_1^2\  \left| {\rm cos} 
\left( 2\sqrt{\pm \frac{\kappa^2 |\sigma|}{3}  } \
                            (t - t_1) \right) \right|  \,. 
\ee 
It describes a Universe which contracts into a singularity in finite time.


\subsection{Radiation dominated Universe}

In the case $k=0$, $\Gamma = 4/3$, $\sigma=0$ 
Eq. (\ref{Heq}) acquires the general relativistic FLRW form, 
\be \label{de_kiirgus_1}
{\dot H} + 2H^2 = 0\,, \qquad \quad
  \qquad \quad \Longrightarrow \qquad \quad
\frac{1}{H} - \frac{1}{H_2}  = 2 (t-t_2)\,,
\ee
where $H_2$, $t_2$ are constants of integration,
$H(t_2) = H_2$. 
The Hubble parameter decreases ($\dot{H} < 0$) and
the scale factor is 
\be \label{a4/3}
a = a_2 \sqrt{2 H_2(t- t_2) +1 } \,,
\ee
where $a_2$ is a  constant of integration, $a(t_2) = a_2$.
From Eqs. (\ref{de_kiirgus_1}) and (\ref{a4/3})  
the Hubble parameter 
can be given in terms of the  scale factor,
\be \label{H^2a^4}
H^2 = \frac{H_2^2 \ a^4_2}{a^4} \,. 
\ee
Since it must coincide with 
the first integral (\ref{H^2rad}) 
the initial value $C$ of the dark radiation density at $t_0$
can be expressed in terms of integration constants,
\be \label{C_and_rho0}
C  + \rho_0 = \frac{3 \  H_2^2 \ a_2^4 }{\kappa^2 \ a_0^4} \,.
\ee

Eq. (\ref{deq}) 
for a  redefined radion field $\chi^2 \equiv 1 - \Psi$ 
takes a simple form,
\be
\ddot{\chi} + 3H \dot{\chi} = 0 \, ,
\ee
and its solution reads
\be \label{chi4/3}
\chi - \chi_{\infty} = - \frac{ g_2}{H_2 \sqrt{2H_2(t - t_2)+1 }} \ , 
\ee
where $\chi_{\infty}$ and $g_2$ are integration constants.
  The constraint equation (\ref{00}) with $k=0$, $V=0$ 
allows us to determine one of them
\be \label{con}
\Psi_{\infty} \equiv 1 - \chi_{\infty}^2 =
  \frac{ \kappa^2}{3H_2^2 a_2^4} \left( \rho_0 + \rho_0^{_B} \chi_0^4\right)
  a_0^4 \ ,
\ee
where $\rho_0$, $\rho_0^B$, $\chi_0^2$, $a_0$ are initial values 
at $t = t_0$ in the
solution of the conservation laws (\ref{ceA}), (\ref{ceB}).
In a realistic cosmological model, 
the radiation dominated 
era has a finite duration $t \in [t_{in}, t_{out}]$. 
Substituting 
$t_2 = t_{in}$ and $H_2 = H_{in}$  into Eqs. (\ref{a4/3}) and (\ref{chi4/3}),
the change in the scalar field can be given in terms of e-folds,
$ N = \ln (a_{out}/a_{in})$, as 
\be \label{deltachi}
\chi_{out} - \chi_{in} = \frac{\dot{\chi}_{in}}{H_{in}}
                  \left(1 - e^{-N}\right) \,.
\ee
If the duration of the radiation dominated era is at least several 
e-folds, the change of the scalar field is approximately determined
only by the values at $t_{in}$. This also means that after a few
e-folds the scalar field remains approximately constant.

The constraint between integration constants  (\ref{con}) can also be
used for putting Eq. (\ref{H^2a^4}) for $H^2$ into a form suitable
for comparison with general relativity,
\be \label{H^2_chi}
H^2 = \frac{\kappa^2}{3 \Psi_{\infty} } \left(1 + 
(1-\Psi_0)^2 \frac{\rho_0^{_B}}{\rho_0}\right) \rho \,.
\ee 
Denoting the Hubble parameter of general relativity by
$H_{_G}$ and taking into account the standard Friedmann equation
$H_{_G}^2 = (8 \pi G /3) \rho$ (here $G$ denotes Newtonian gravitational 
constant which in the context of scalar-tensor theory
depends on the asymptotic scalar field, see Subsect. 6.1)
the  speed-up factor $\xi$ for radiation dominated 
Universe turns out to be constant,
\be  \label{speed}
  \xi^2 \equiv \frac{H^2}{H_{_G}^2} = 
\frac{\kappa^2}{8 \pi G \Psi_{\infty}}
  \left(1 + (1- \Psi_0)^2 \ \frac{\rho_0^{_B}}{\rho_0}\right) \,.
\ee 
In terms of the initial values of radiative energy densities $\rho_0$,
$C$, it takes a simple form
\be \label{speedC}
\xi^2 = \frac{\kappa^2}{8 \pi G} \left(1 + \frac{C}{\rho_0}\right) \,.
\ee
Combining the last two equations, one may also express
\be
C = \frac{\rho_0(1-\Psi_{\infty}) + \rho_0^{_B}(1-\Psi_0)^2}{\Psi_{\infty}} \,.
\label{C_and_constants}
\ee
This tells that having dark radiation contributing
towards acceleration, i.e., $C<0$, is only consistent with negative 
energy density for B-brane radiation,
$\rho_0^{_B} < - \frac{1-\Psi_{\infty}}{(1-\Psi_0)^2}\rho_0$.
Yet, since the RHS of Eq. (\ref{C_and_rho0}) is positive, it is clear that the 
net contribution of radiation and dark radiation to acceleration is always
negative.
\label{rad_solutions}


\subsection{Dust dominated Universe}

Integrating the first integral 
(\ref{H^2}) with $k =0$, $\sigma=0$, $\Gamma = 1$
leads to a cubic equation for the scale factor which 
has one real and two complex solutions. 
The real solution reads 
\beq
a  &=& a_3 \ \left (\frac{1}{2}  \  b^{\frac{1}{3}} + 
2 \frac{C^2}{\rho_0^{2}} \ b^{-\frac{1}{3}} + 
\frac{C}{\rho_0} \right)  \ ,
\eeq
where $t_3$, $a_3$ are  integration constants and the
dimensionless quantity $b$ can be written as 
\beq
b  &=& \rho_0^{-3} \ \left (3 \kappa^2 \rho_{0}^4 \ (t - t_3)^2  - 
8 C^3 
+ \sqrt{9 \kappa^4 \rho_{0}^8 \ (t - t_3)^4 - 48 C^3 \kappa^2 
\rho_{0}^4 \ (t - t_3)^2} \right) \ .
\eeq
The meaning of $t_3$ is clarified by the relation
$a(t_3) =a_3 \,\frac{C}{\rho_0}$, which also implies
$C \geq 0$. Note that in the case of vanishing dark radiation
there is a singularity at the moment $t_3$. 
If $C \not= 0$ and we take $t_0$ to be the present time, 
then $t_3 < t_0$, if
at present $\frac{C}{\rho_0}< 1$, and $t_3 > t_0$, if
$\frac{C}{\rho_0}> 1$.

The Hubble parameter as a function of the cosmic time reads 
\beq
H= \sqrt{\frac{ \kappa^2 \rho_0}{3}} \frac{2\left( 1 - 4 \ 
(\frac{C}{\rho_0})^2 b^{-\frac{2}{3}}\right)}
{\sqrt{3 \kappa^2 \rho_0 \ (t - t_3)^2 - 16 \frac{C^3}{\rho_0^3}} \ 
\left( 1 + 2  (\frac{C}{\rho_0}\ b^{-\frac{1}{3}} )+ 
4 \ (\frac{C}{\rho_0})^2 \ b^{-\frac{2}{3}} \right)} \ .
\eeq
In the case of vanishing dark radiation $C=0$ we get the FLRW
value for the Hubble parameter $H = \frac{2}{3} (t-t_3)^{-1}$.


\subsection{Universe with cosmological constant and radiation}

Eq. (\ref{Heq}) with $k=0$, $\sigma \not= 0$ does not
depend on  radiation matter with $\Gamma = \frac{4}{3}$, 
so its solutions coincide with those for cosmological constant
dominated Universe (\ref{H_G4/3_k0}). In particular, Eq. (\ref{fi})
holds and the late time Universe is dominated by the cosmological
constant $\sigma$ alone.

For getting an equation for the radion field $\chi^2 \equiv 1 - \Psi$
let us solve the corresponding constraint equation (\ref{00})
as an algebraic equation for $\dot{\chi}$
\be
\dot{\chi} = -H\chi \pm \sqrt{H^2 - \frac{\kappa^2}{3} \left(\rho +
\sigma + \chi^4(\rho^{_B} +\sigma^{_B})\right)}\,.
\ee 
Taking into account Eq. (\ref{fi}) for $H^2$ and introducing the 
conformal time $d\eta = a^{-1}dt$ we get
\be
\frac{d}{d\eta} \left(a \chi \right) =
\pm \frac{\kappa}{\sqrt{3}} \sqrt{(C - \rho_0) a_0^4
+(\sigma^{_B} + \rho_0^{_B}) (a \chi)^4} \,,
\ee
which in principle can be integrated in terms of elliptic functions.


\section{Evolution of the scalar field}


In the previous section it was possible to find analytic solutions for
the scale factor since the equations of motion (\ref{00})--(\ref{deq}) 
combined to yield an equation for $H$ decoupled from the 
scalar field $\Psi$ and B-brane matter. 
In fact, it is also possible to derive an independent equation for $\Psi$ if
we retain only one type of source (cosmological constants or matter) 
similar on both branes ($\Gamma^{^B} = \Gamma$) and take the Universe 
to be flat ($k=0$). Then, after introducing a new time variable \cite{sa1}
\be \label{p time}
dp=h_c \, dt , \quad h_c = H + \frac{\dot{\Psi}}{2\Psi} , \quad
\frac{df}{dp}=f' \,,
\ee
the equations (\ref{00})--(\ref{deq})
together with (\ref{rhoAB}) give a decoupled ``master equation'' for $\Psi$,
\beq \label{master equation}
8 (1-\Psi)\frac{\Psi''}{\Psi} - 3(2-\Gamma)\left( \frac{\Psi'}{\Psi}
\right)^3
-2 \Bigl( (4-6\Psi)- (4-3\Gamma)(1-\Psi) W(\Psi)
\Bigr)\left(\frac{\Psi'}{\Psi}\right)^2 && \nonumber \\
+ 12 (2-\Gamma)(1-\Psi)\frac{\Psi'}{\Psi} - 8(4-3\Gamma)(1-\Psi)^2 W(\Psi)
=& 0, &
\eeq
where for the cosmological constants ($\Gamma=0$)
\be
W(\Psi) = \frac{1+(1-\Psi) \, \frac{\sigma^{_B}}{\sigma}}{1+(1-\Psi)^2 \,
\frac{\sigma^{_B}}{\sigma}} \; ,
\ee
and for matter ($\Gamma>0$)
\be
W(\Psi) = \frac{ 1+(1-\Psi) \, \frac{\rho^{_B}_0}{\rho_0} \left(
\frac{1-\Psi}{1-\Psi_0} \right)^{-\frac{3}{2}\Gamma}}
{1+(1-\Psi)^2 \, \frac{\rho^{_B}_0}{\rho_0} \left( \frac{1-\Psi}{1-\Psi_0}
\right)^{-\frac{3}{2}\Gamma}} \, .
\ee
To solve this master equation analytically is still 
a formidable task, but in order to get a qualitative picture of the evolution 
of $\Psi$ we can study Eq. (\ref{master equation}) as dynamical system.

\subsection{Cosmological constant dominated Universe}

By defining the variables $x \equiv \Psi$, $y \equiv \Psi'$ we can rewrite
the equation (\ref{master equation}) for $\Gamma=0$ as a dynamical system:
\beq
x' & = & y \nonumber \\
y' & = & \frac{3 y^3}{4 x^2 (1-x)} + \frac{ \left( (4-6x)-4(1-x)W(x) \right)
y^2}{4 x (1-x)}
-3y + 4x(1-x)W(x) \, .
\label{sigma_dynsys}
\eeq
The physically allowed regions of the phase space ($x, y$) 
are constrained by 
the Friedmann equation (\ref{00}), which in terms of the new variables
turns out to be
\be \label{Fr_cc}
h_c^2 = \frac{\kappa^2 \left( \sigma + (1-x)^2 \sigma^{_B} \right)}
{3x \left( 1- \frac{{y}^2}{4 x^2(1-x)} \right)} \; .
\ee
If we let $\sigma$ and $\sigma^{_B}$ to take positive and negative values,
there are four possibilities, summarized in Table \ref{sigma_Friedmann_allow}.
Note, Eq. (\ref{Fr_cc}) also assures that the time reparametrization 
(\ref{p time}) is acceptable --- within the borders 
of the allowed phase space $t$-time and $p$-time always run in the same
direction.
\begin{table}[t] 
\begin{center}
\begin{tabular}{lccccc}
& Cosmological constants & \qquad & Interval of $\Psi$ & \quad & Allowed region \vspace{2mm}\\
\hline \\
1) & $\sigma\geq0$, $\sigma + \sigma^{_B}\geq0$ && $0 \leq x \leq 1$ &&
$|y| \leq |2x \sqrt{1-x}|$ \vspace{2mm}\\
2) & $\sigma \geq0$, $\sigma + \sigma^{_B}\leq0$ && $0 \leq x \leq
1-\sqrt{-\frac{\sigma}{\sigma^{_B}}}$ && $|y| \geq |2x \sqrt{1-x}|$
\\
  & && $1-\sqrt{-\frac{\sigma}{\sigma^{_B}}} \leq x \leq 1$ && $|y| \leq
|2x \sqrt{1-x}|$ \vspace{2mm}\\
3) & $\sigma \leq0$, $\sigma + \sigma^{_B}\geq0$ && $0 \leq x \leq
1-\sqrt{-\frac{\sigma}{\sigma^{_B}}}$ && $|y| \leq |2x \sqrt{1-x}|$
\\
  & && $1-\sqrt{-\frac{\sigma}{\sigma^{_B}}} \leq x \leq 1$ && $|y| \geq
|2x \sqrt{1-x}|$ \vspace{2mm}\\
4) & $\sigma\leq0$, $\sigma + \sigma^{_B}\leq0$ && $0 \leq x \leq 1$ &&
$|y| \geq |2x \sqrt{1-x}|$ \vspace{2mm}\\
\hline
\end{tabular}
\end{center}
\caption{Regions on the $(x=\Psi,y=\Psi')$ plane allowed for dynamics
with cosmological constants.}
\label{sigma_Friedmann_allow}
\end{table}

In order to understand how to read the phase portraits correctly,
it is instructive to see how the velocities $\dot{\Psi}$
get mapped to $\Psi'$ by 
\be
\dot{\Psi} = \frac{d\Psi}{dp} \frac{dp}{dt} = y h_c \,.
\ee
It turns out that the boundary
\be
\rm{B}_{\pm}: \ \ y = \pm 2x \sqrt{1-x} 
\ee
corresponds to $\dot{\Psi} = \pm \infty$,
while $y=0$ corresponds to $\dot{\Psi} = 0$. The outer reaches,
$y= \pm \infty$, which are only relevant when 
$\sigma + (1-x)^2 \sigma^{_B} <0$, translate into 
\be
\dot{\Psi}|_{y=\pm \infty} = 
\pm \kappa \sqrt{-\frac{4}{3}x(1-x)(\sigma + (1-x)^2 \sigma^{_B})},
\label{y_inf}
\ee
thus we get a minimal bound on the speed of the relative brane motion.
At the borders $x=0$ and $x=1$ we have $\dot{\Psi} = 0$
and the same holds for the points $(x=0,y=0)$
and $(x=1,y=0)$ as well. This tells us immediately that the brane 
collision ($\Psi=0$) can take place only extremely softly at vanishing 
relative speed ($\dot{\Psi} \rightarrow 0^{-}$), and also when the branes
approach infinite separation ($\Psi=1$), they essentially stop moving with
respect to each other ($\dot{\Psi} \rightarrow 0^{+}$).
At the crossing points 
\be \label{crossing_points}
\rm{C}_{\pm}= \left( x=1-\sqrt{-\frac{\sigma}{\sigma^{_B}}},\
y=\pm 2
\left(1-\sqrt{-\frac{\sigma}{\sigma^{_B}}}\right) \sqrt[4]{-\frac{\sigma}{\sigma^{_B}}}
\right)
\ee
the ratio $\frac{dp}{dt}$ becomes undefined, meaning that
with our redefinition of time (\ref{p time})
at these values of $\Psi$ all velocities $\dot{\Psi}$ get mapped to a single
respective velocity $\Psi'$.

\begin{figure}[t]
\begin{center}
\includegraphics[width=7cm,height=7cm,angle=-90]{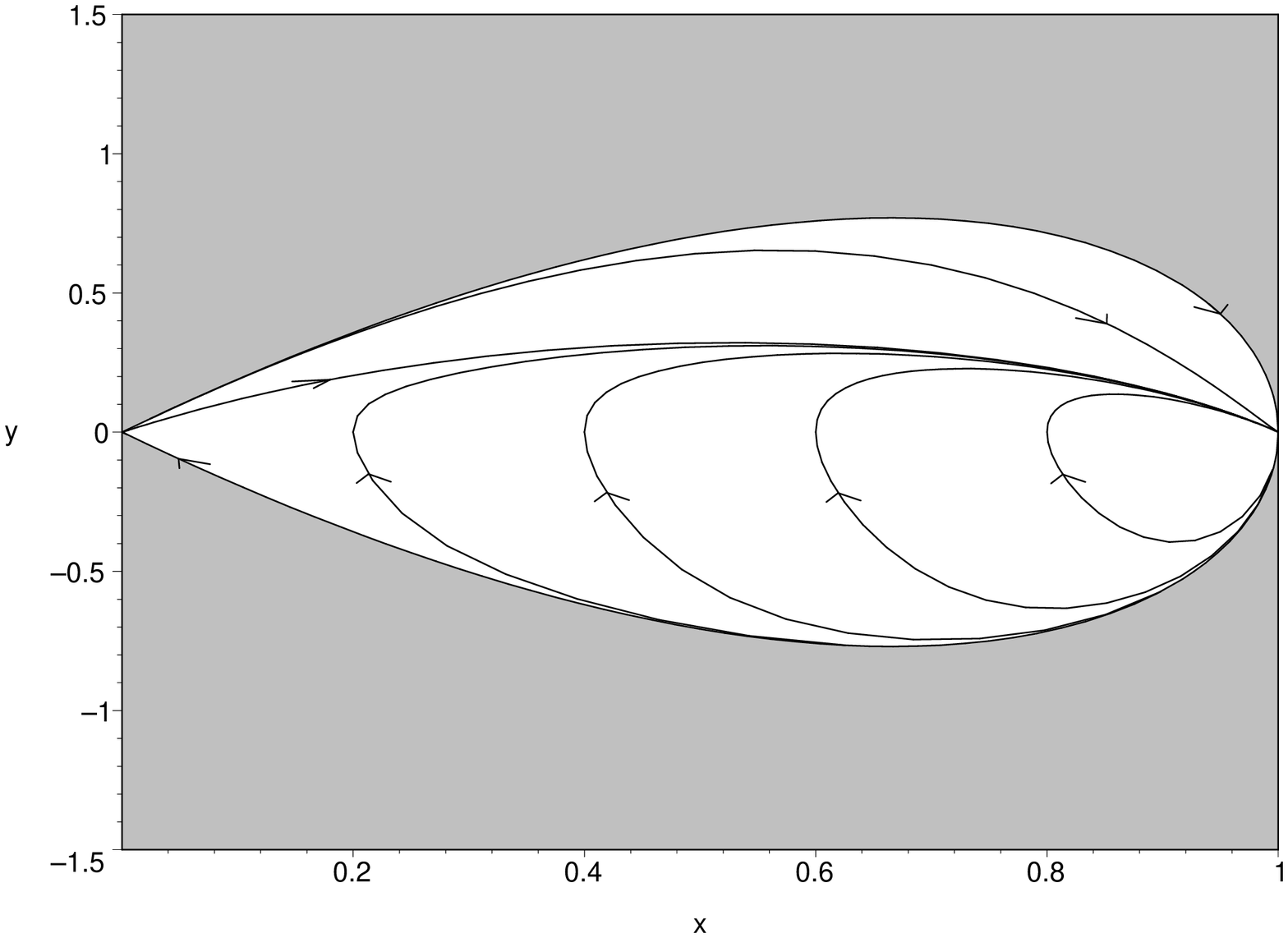}
\hspace{0.5cm}
\includegraphics[width=7cm,height=7cm,angle=-90]{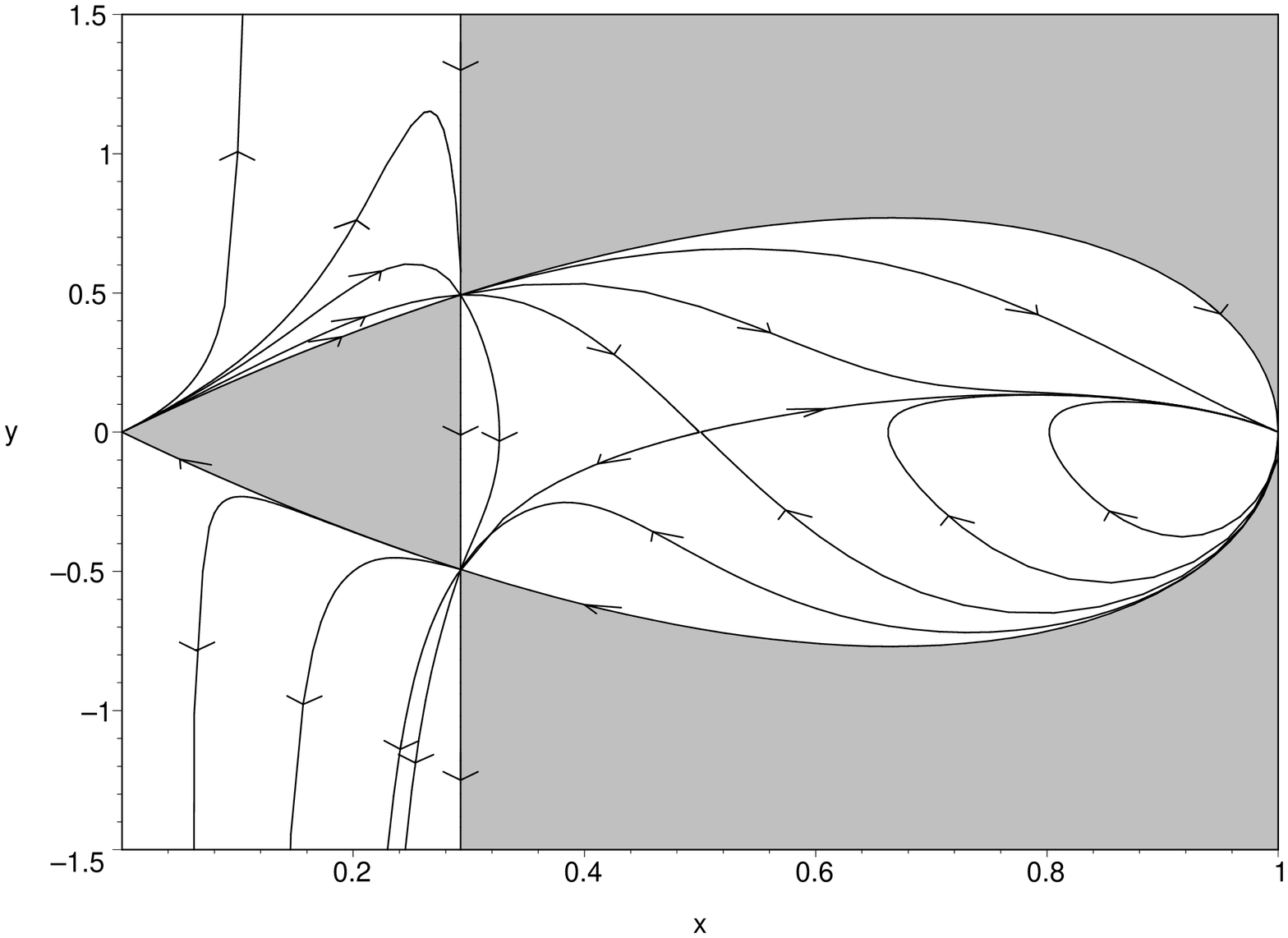}
\vspace{0.5cm}
\includegraphics[width=7cm,height=7cm,angle=-90]{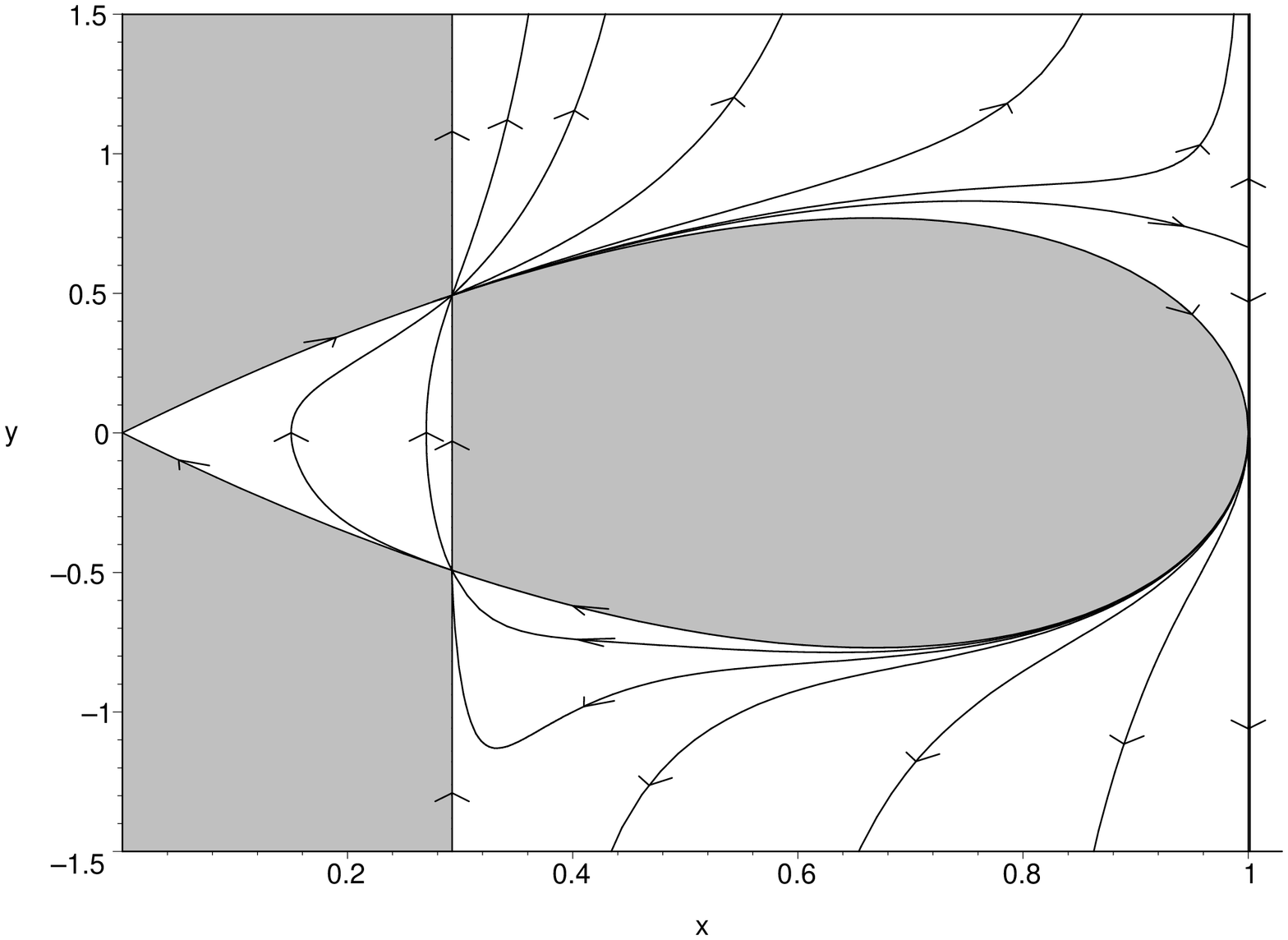}
\hspace{0.5cm}
\includegraphics[width=7cm,height=7cm,angle=-90]{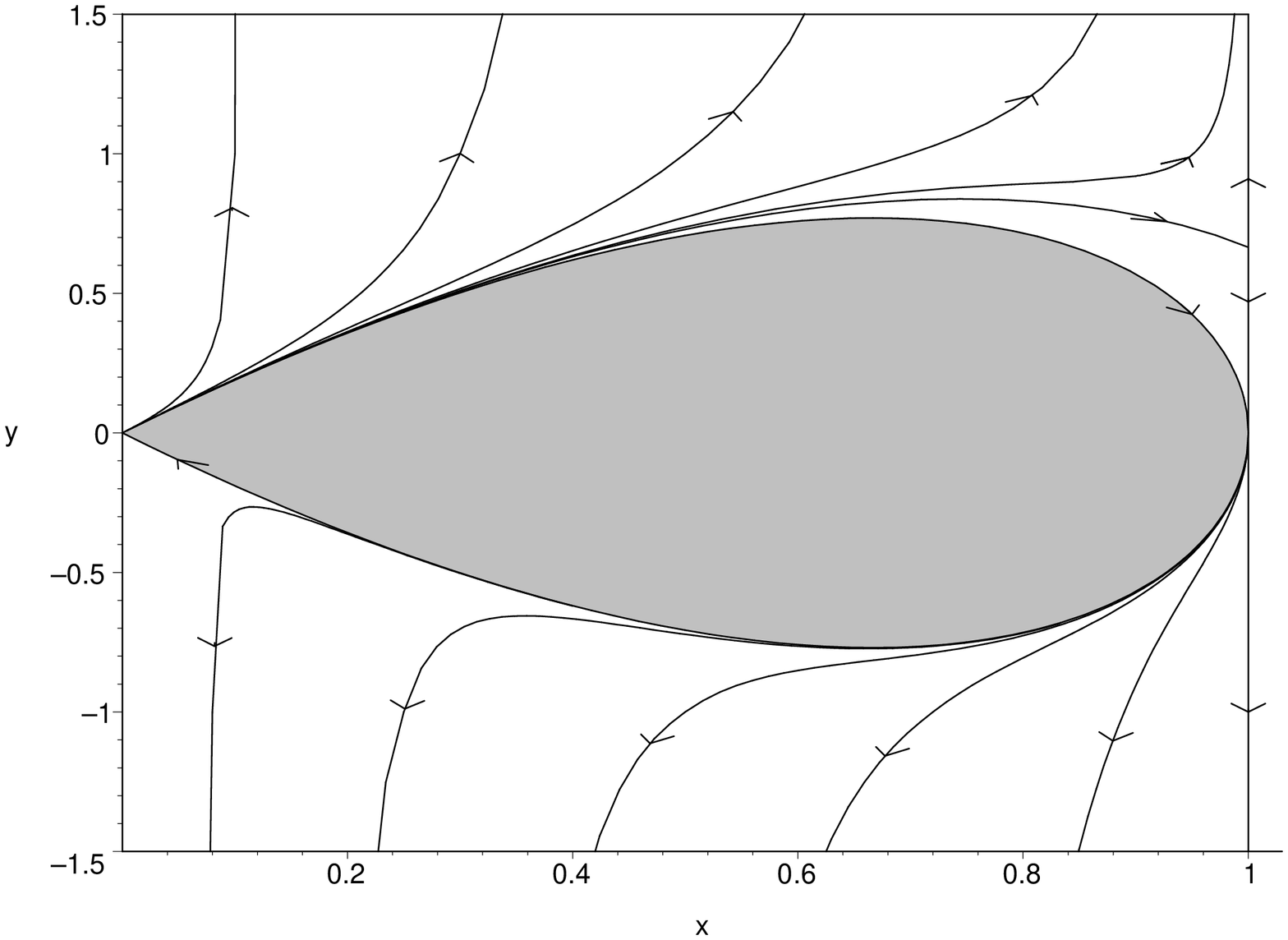}
\end{center}
\caption{Phase portraits ($x=\Psi(p)$, $y=\Psi'(p)$) for cosmological 
constants on both branes: 
1) $\sigma=1, \sigma^{_B}=-0.5$ (top, left), 
2) $\sigma=0.5, \sigma^{_B}=-1$ (top, right), 
3) $\sigma=-0.5, \sigma^{_B}=1$ (bottom, left), 
4) $\sigma=-1, \sigma^{_B}=0.5$ (bottom, right).}
\label{plots}
\end{figure}

The system (\ref{sigma_dynsys}) possesses the following fixed points:
\bi
\item $\rm{P}_1 = (x=0, y=0)$, a saddle point with repulsive and attractive
eigenvectors tangential
to the upper and lower boundaries  $\rm{B}_{\pm}$, respectively;
\item $\rm{P}_2 = (x=1, y=0)$, a (complex) spiralling attractor, but here 
the trajectories have to obey also the boundaries of the allowed region;
\ei
and in the case $\sigma \geq0$, $\sigma + \sigma^{_B}\leq0$ there is also
\bi
\item $\rm{P}_3 = (x=1+\frac{\sigma}{\sigma^{_B}}, y=0)$, a saddle point
(having de Sitter expansion for the scale factor \cite{kss,iii}).
\ei
The limit of general relativity, ${\dot \Psi} =0$,  $\Psi = 1$, 
necessarily holds for all trajectories drawn to the right boundary $x=1$. 
The phase portraits in the four cases are depicted in Figure \ref{plots},
let us consider these one by one.

In the case 1) of $\sigma\geq0$, $\sigma + \sigma^{_B}\geq0$ 
(top left on Figure \ref{plots}) 
the allowed phase space is restricted into one bounded region, with
$\dot{\Psi} = \pm \infty$ getting mapped to the boundaries $\rm{B}_{\pm}$ 
and all the finite values being in the middle. 
Trajectories can either start 
near $\rm{P}_1$, where the two branes very close to each other
slowly start to move apart, 
or near $\rm{P}_2$, where the very far away branes slowly begin to 
move towards each other. The latter class of trajectories will eventually
turn around, meaning these branes will never meet 
but travel back towards infinite separation.
Hence all the trajectories are collected by the attractor $\rm{P}_2$, which 
corresponds to the limit of general relativity. 
(Of course, by definition the two fixed points provide 
two stationary trajectories themselves, i.e., two overlapping branes and 
two infinitely far away branes would always stay so.)

In the case 2) with $\sigma \geq0$, $\sigma + \sigma^{_B}\leq0$
(top right on Figure \ref{plots}) 
the allowed phase space consists of three areas, connected via two 
crossing points $\rm{C}_{\pm}$. 
Notice that in the left areas $\dot{\Psi} = \pm \infty$ gets mapped
to the boundaries $\rm{B}_{\pm}$, while 
$y=\pm \infty$ corresponds to a finite velocity (\ref{y_inf}).
Again, the trajectories can either start close to $\rm{P}_1$ 
or close to $\rm{P}_2$. 
The former trajectories explore the upper left allowed phase
space region, pass through the upper crossing point $\rm{C}_{+}$
into the right allowed region 
and then proceed either to the fixed point $\rm{P}_2$, or pass 
though the lower crossing point $\rm{C}_{-}$ into the lower
left allowed region to run into $(x=0,y=- \infty )$, or in the marginal case
terminate at the saddle point $\rm{P}_3$.
These correspond to branes either starting very close to each 
other and moving apart towards infinity, or turning back and colliding, 
or, marginally, stopping 
at a finite distance. The other class of trajectories beginning near 
$\rm{P}_2$ have the same possibilities, i.e., either they
 go back to $\rm{P}_2$, 
or pass through the crossing point $\rm{C}_-$ into the lower left region 
to run into $(x=0,y=- \infty )$, and there is also a marginal one 
that terminates 
at the saddle point $\rm{P}_3$. These correspond to branes starting to move
closer to each other from a large distance, and then either separate again, 
or eventually collide, or in the marginal case come to a standstill
at certain finite distance. 

In the case 3) with $\sigma \leq0$, $\sigma + \sigma^{_B}\geq0$ 
(bottom left on Figure \ref{plots}) the allowed regions are 
an inverse of the previous case. Now the common trajectories may only start 
near $\rm{P}_2$, explore the lower right region, necessarily pass through $\rm{C}_-$, 
visit the left allowed region, pass through $\rm{C}_+$, and then either run
to $\rm{P}_2$, or to $(x=1,y= \infty )$, or, marginally, to $(x=1,y=2/3)$. 
These correspond to far apart branes approaching each other to at least 
a certain distance, and then separating again towards infinity.
All these trajectories will 
gently deliver the limit of general relativity (as they get to $x=1$). 
The distinction where they end on the $y$ axis is a coordinate
artifact of time $p$, simply
meaning how ``fast'' their speed measured in time $t$, i.e., $\dot{\Psi}$ 
tends to zero.

In the case 4) with $\sigma\leq0$, $\sigma + \sigma^{_B}\leq0$ 
(bottom right on Figure \ref{plots}) the allowed part of the phase space
consists of two separate regions. The trajectories can either begin
near $\rm{P}_2$ and go to $(x=0,y= -\infty )$, or start near $\rm{P}_1$ 
and end up at $\rm{P}_2$, $(x=1,y=2/3)$, or $(x=1,y= \infty )$. The
former trajectories describe branes coming closer from far apart to a collision, 
while the latter trajectories describe branes starting very close to
each other and moving apart towards infinity, leading
to the limit of general relativity.

Some qualitative features of the radion dynamics in the cosmological 
constants case were inferred previously by Kanno, Soda, and Sasaki 
\cite{kss} 
by considering the Einstein frame radion potential, 
and by Webster and Davis \cite{webster_davis_MSA} in the Jordan frame 
moduli space approximation. Both papers also point out that the 
A-brane Hubble parameter does not diverge at the event of brane collision, thus 
leading to the picture that branes could pass through each other 
with possible interesting cosmological consequences. 
Here we observe that the relative brane velocity in the cosmic time 
slows down and vanishes at the collision. 
However, this result must be taken with considerable caution, since one
expects the effective action to receive corrections near the collision 
thus changing the behavior of the system \cite{close_branes}.

\subsection{Radiation dominated Universe}
\label{rad_dom_dyn_sys}

\begin{table}[t] 
\begin{center}
\begin{tabular}{lccccc}
& Radiation & \qquad & Interval of $\Psi$ & \quad & Allowed region \vspace{2mm}\\
\hline \vspace{-2mm} \\ 
1) & $\rho^{_B} \geq 0$ && $0 \leq x \leq 1$ &&
$|y| \leq |2x \sqrt{1-x}|$ \vspace{2mm}\\
1a) & $\rho^{_B} <0, \rho_0 + (1-\Psi_0)^2 \rho^{_B}_0 \geq 0$ && $0 \leq x \leq
1$ && $|y| \leq |2x \sqrt{1-x}|$
\vspace{2mm}\\
2) & $\rho^{_B} <0, \rho_0 + (1-\Psi_0)^2 \rho^{_B}_0 \leq 0$ && $0 \leq x \leq
1$ && $|y| \geq |2x \sqrt{1-x}|$ \vspace{2mm}
\\
\hline \\
\vspace{0mm}\\
& Dust & \qquad & Interval of $\Psi$ & \quad & Allowed region \vspace{2mm}\\
\hline \vspace{-2mm}\\
1) & $\rho^{_B} \geq 0$ && $0 \leq x \leq 1$ &&
$|y| \leq |2x \sqrt{1-x}|$ \vspace{2mm}\\
1a) & $\rho^{_B} < 0$, $\rho_0 + (1-\Psi_0)^{3/2} \rho^{_B}_0 \geq 0$ 
&& $0 \leq x \leq 1$ && $|y| \leq |2x \sqrt{1-x}|$
\vspace{2mm}\\
2) & $\rho^{_B} <0$, $\rho_0 + (1-\Psi_0)^{3/2} \rho^{_B}_0 \leq 0$ 
&& $0 \leq x \leq 1 - \frac{\rho_0^2}{ {\rho^{_B}_0}^2 (1-\Psi_0)^{3}} $ 
&& $|y| \geq |2x \sqrt{1-x}|$
\\
  & && $1 - \frac{\rho_0^2}{ {\rho^{_B}_0}^2 (1-\Psi_0)^{3}} \leq x \leq 1$ 
&& $|y| \leq
|2x \sqrt{1-x}|$ \vspace{2mm}\\
\hline
\end{tabular}
\end{center}
\caption{Regions on the $(x=\Psi,y=\Psi')$ plane allowed for dynamics 
with radiation and dust (assuming $\rho > 0$).}
\label{matter_Friedmann_allow}
\end{table}

For radiation on both branes ($\Gamma = \Gamma^{^B}=4/3$) the dynamical 
system of Eq. (\ref{master equation}) reduces to
\beq \label{rad_sys}
x' & = & y \nonumber \\
y' & = & \frac{y^3}{4 x^2 (1-x)} + \frac{(4-6x)y^2}{4 x (1-x)} -y \,\, .
\eeq
The Friedmann constraint is similar to the cosmological constants case, 
\be
h_c^2 = \frac{\kappa^2 \left( \rho + (1-x)^2 \rho^{_B} \right)}
{3x \left( 1- \frac{{y}^2}{4 x^2(1-x)} \right)} =
\frac{\kappa^2 \left( \rho_0 + (1-\Psi_0)^2 \rho^{_B}_0 \right)}
{3x \left( 1- \frac{{y}^2}{4x^2(1-x)} \right)} 
\left( \frac{a}{a_0} \right)^{-4} \; ,
\ee
where in the last step we used Eqs. (\ref{ceA}), (\ref{rhoAB}) 
to express $\rho$ and $\rho^{_B}$. The allowed regions are 
listed in Table \ref{matter_Friedmann_allow}, including the 
exotic possibility of $\rho^{_B}<0$. 
The discussion of how $\dot{\Psi}$ is mapped to $\Psi'$
follows along the same lines as before yielding analogous results,
only the outer reaches, $y= \pm \infty$, relevant when 
$\rho_0 + (1-\Psi_0)^2 \rho_0^{_B} <0$, translate into 
\be
\dot{\Psi}|_{y=\pm \infty} = \pm \kappa \sqrt{-\frac{4}{3}x(1-x)
(\rho_0 + (1-\Psi_0)^2 \rho_0^{_B})} \left( \frac{a}{a_0} \right)^{-2} \,.
\label{y_inf_rad}
\ee
The limit of general relativity is still approached on the line $x=1$.
However, the main new feature is that the system (\ref{rad_sys}) does not have 
any fixed points, since the force term has cancelled. 

\begin{figure}[t]
\begin{center}
\includegraphics[width=7cm,height=7cm,angle=-90]{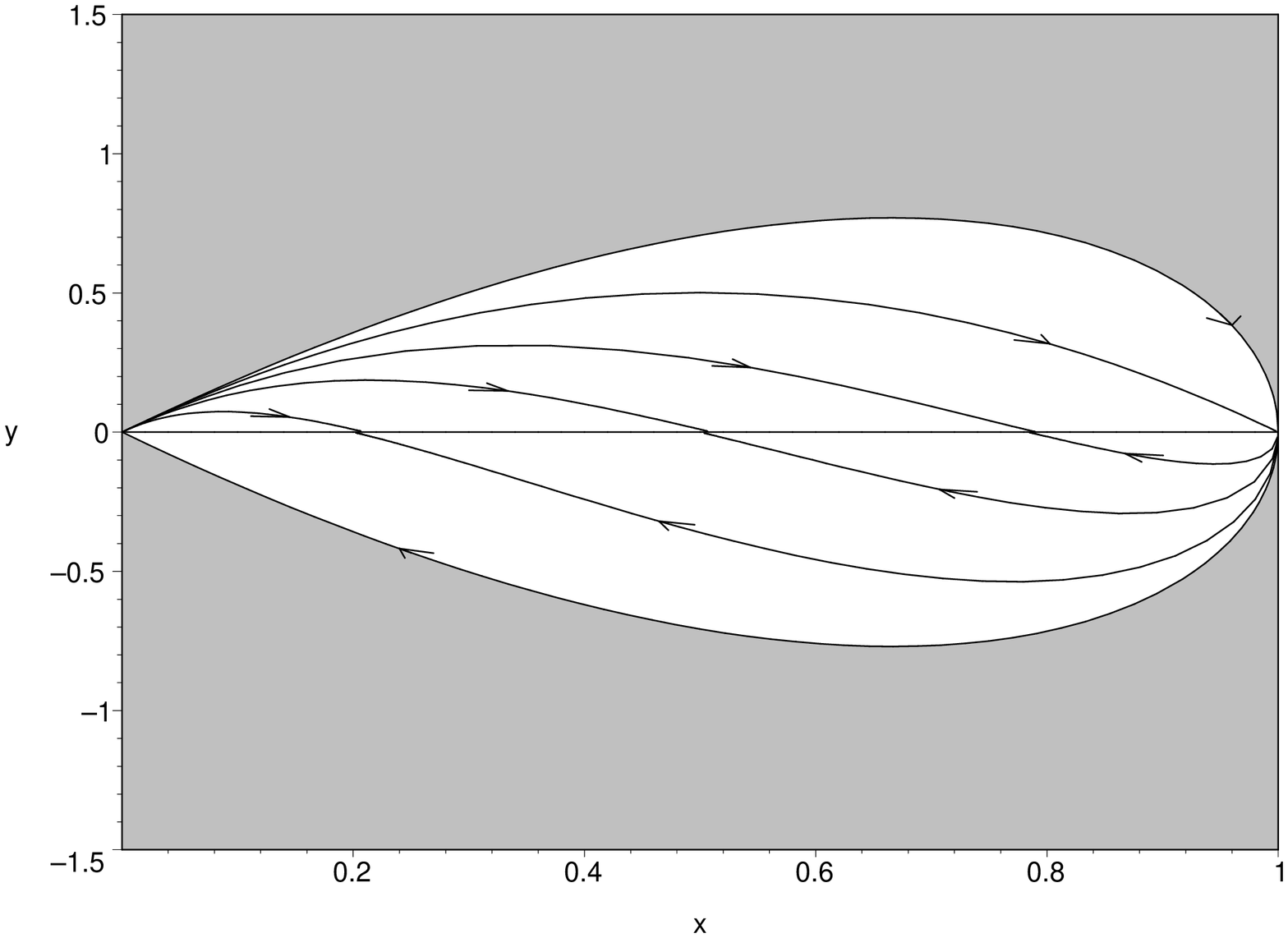}
\hspace{0.5cm}
\includegraphics[width=7cm,height=7cm,angle=-90]{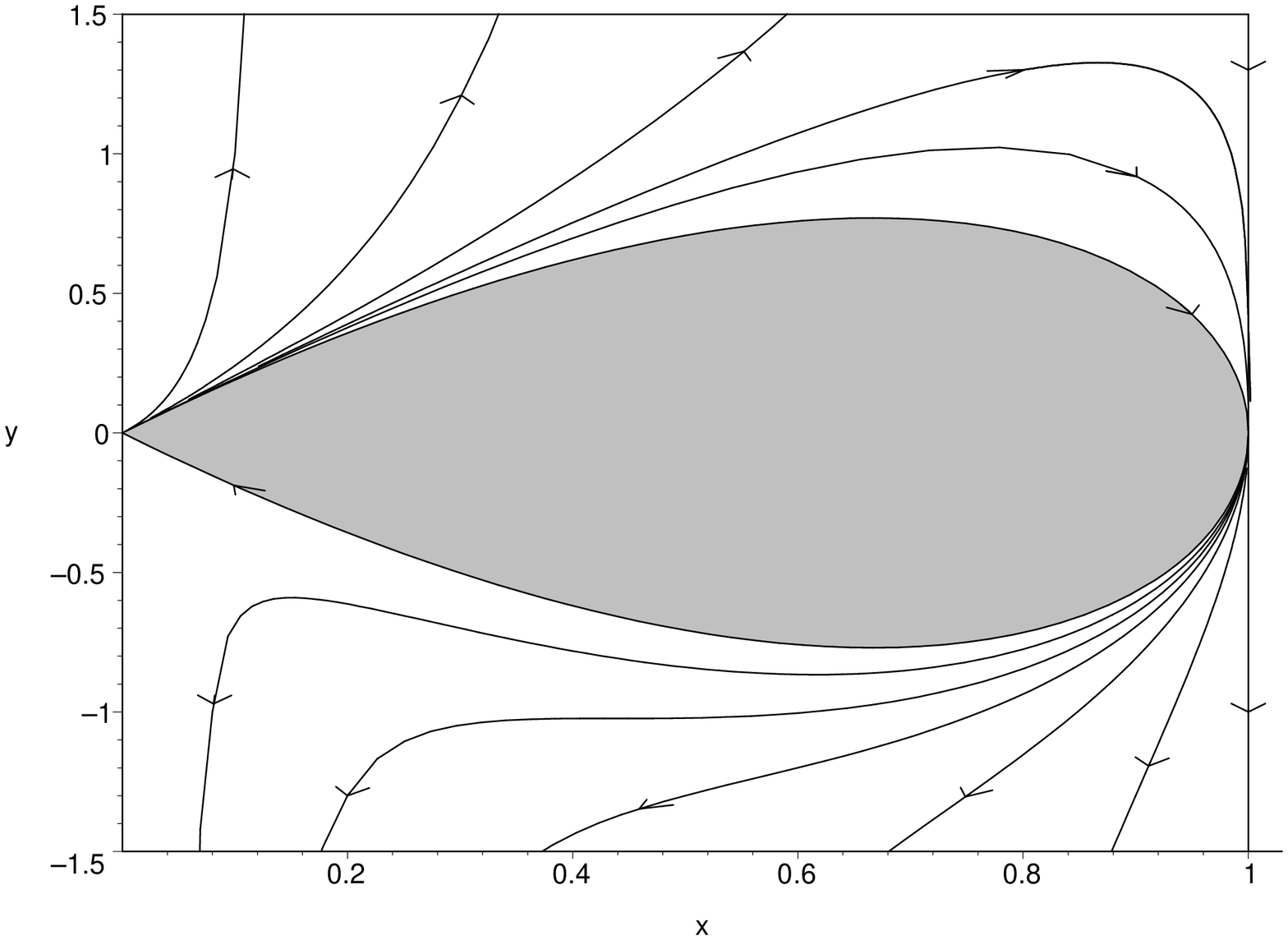}
\vspace{0.5cm}
\includegraphics[width=7cm,height=7cm,angle=-90]{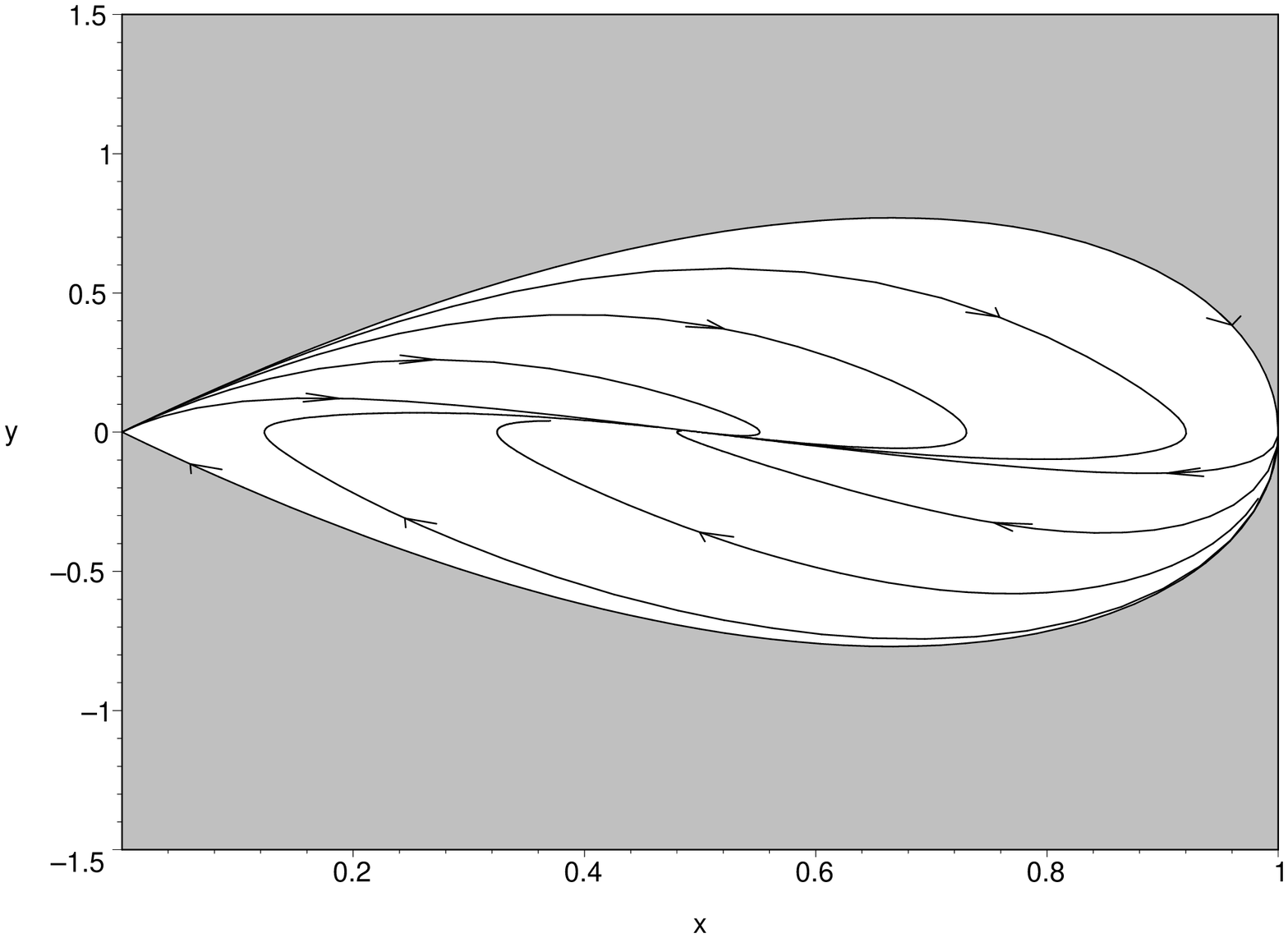}
\hspace{0.5cm}
\includegraphics[width=7cm,height=7cm,angle=-90]{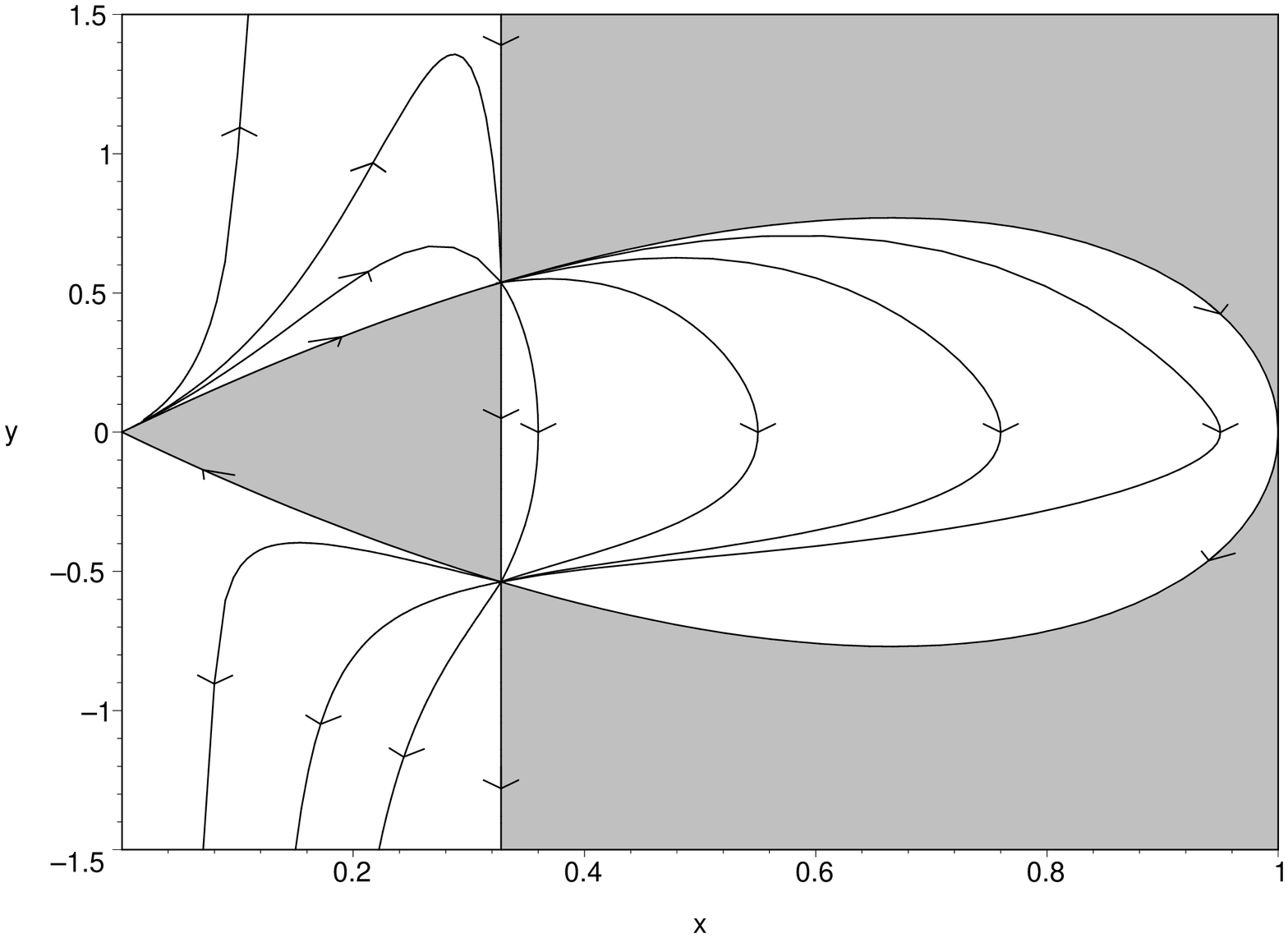}
\end{center}
\caption{Phase portraits ($x=\Psi(p)$, $y=\Psi'(p)$) for radiation: 
1) $\rho=1, \rho^{_B}=1, \Psi_0=0.1$ (top, left), 
2) $\rho_0=0.5, \rho^{_B}_0=-1, \Psi_0=0.1$ (top, right), 
and for dust: 
1a) $\rho_0=0.5, \rho^{_B}_0=-1, \Psi_0=0.5$ (bottom, left), 
2) $\rho_0=0.7, \rho^{_B}_0=-1, \Psi_0=0.1$ (bottom, right).}
\label{plots_rad}
\end{figure}

In the case 1) with $\rho$ and $\rho^{_B}$ both positive, and also in the
more exotic case 1a)
with $\rho^{_B}<0$, but $\rho_0 \geq (1-\Psi_0)^2 \rho^{_B}_0$ 
the allowed phase space covers a single region 
(Figure \ref{plots_rad} top left).
The trajectories can either start near $(x=0,y=0)$ or $(x=1,y=0)$, 
but all of them end somewhere on the $y=0$ line. 
This corresponds
to close branes separating or far away branes approaching each other, but 
eventually all of them slowing down and stopping at some finite distance. 
The whole dynamics is ruled by friction,
which is also in accord with our observation in Subsection
\ref{rad_solutions} that after a few e-folds of expansion the scalar 
field remains approximately constant. 

In the other exotic case 2) with $\rho^{_B}<0$ and 
$\rho_0 \leq -(1-\Psi_0)^2 \rho^{_B}_0$ there are two allowed regions 
(Figure \ref{plots_rad} top right).
The trajectories beginning near $(x=0,y=0)$ run through the upper region
into $(x=1,y=0)$, meaning
close branes separating towards infinity and reaching the limit of 
general relativity. 
The trajectories starting
near $(x=1,y=0)$ run through the lower region into $(x=0,y=-\infty)$, 
meaning far away branes coming to collide. 


\subsection{Dust dominated Universe}

Taking both branes to be dominated by dust ($\Gamma = \Gamma^{^B}=1$) 
the dynamical system is given by
\beq
x' & = & y  \nonumber \\
y' & = & \frac{3 y^3}{8 x^2 (1-x)} + \frac{ \left( (4-6x)-(1-x)W(x) \right)
y^2}{4 x (1-x)}
-\frac{3 y}{2} + x (1-x) W(x) \,.
\label{dust_dynsys}
\eeq
The allowed regions are determined by the Friedmann constraint,
\be
h_c^2 = \frac{\kappa^2 \left( \rho + (1-x)^2 \rho^{_B} \right)}
{3x \left( 1- \frac{{y}^2}{4 x^2(1-x)} \right)} =
\frac{\kappa^2 \left( \rho_0 + (1-x)^{1/2}(1-\Psi_0)^{3/2} \rho^{_B}_0 
\right)}
{3x \left( 1- \frac{{y}^2}{4x^2(1-x)} \right)} 
\left( \frac{a}{a_0} \right)^{-3} \; ,
\ee
where again in the last step we used Eqs. (\ref{ceA}), (\ref{rhoAB}) 
to express $\rho$ and $\rho^{_B}$. The regions available for
the dynamics are listed in Table \ref{matter_Friedmann_allow}. Translating
$\dot{\Psi}$ into $\Psi'$ proceeds with analogous results, only
only the outer reaches, $y= \pm \infty$, relevant when 
$\rho_0 + (1-\Psi_0)^{3/2} \rho_0^{_B} <0$, correspond to 
\be
\dot{\Psi}|_{y=\pm \infty} = \pm \kappa \sqrt{-\frac{4}{3}x(1-x)
\left( \rho_0 + (1-x)^{1/2}(1-\Psi_0)^{3/2} \rho_0^{_B} \right)} 
\left( \frac{a}{a_0} \right)^{-3/2} \,.
\label{y_inf_dust}
\ee
The limit of general relativity is $x=1$ as in the previous cases.
The system (\ref{dust_dynsys}) is endowed with three fixed points:
\bi
\item $\rm{P}_1 = (x=0, y=0)$, a saddle point with repulsive and attractive
eigenvectors tangential
to the upper and lower boundaries  $\rm{B}_{\pm}$, respectively;
\item $\rm{P}_2 = (x=1, y=0)$, a (complex) spiralling attractor for 
$\rho^{_B}_0 \geq 0$, but a saddle point for $\rho^{_B}_0 < 0$, notice
that around here 
the trajectories have to obey also the boundaries of the allowed region;
\ei
and in the case $\rho^{_B}_0 < 0$, $\rho_0 + \rho^{_B}_0 (1-\Psi_0)^{3/2} \geq 0$ 
there exists also
\bi
\item $\rm{P}_3 = (x=1-\frac{{\rho^{_B}_0}^2 (1-\Psi_0)^3}{\rho^2_0}, y=0)$, an attractor.
\ei

In the case 1) with $\rho_0 > 0$, $\rho^{_B}_0 > 0$ the phase portrait is 
qualitatively identical to case 1) with cosmological constants 
(Figure \ref{plots} top right). 
The trajectories can start either at $\rm{P}_1$ or $\rm{P}_2$, but they
are all attracted to $\rm{P}_2$, approaching the limit of general relativity.
The case 1a) with $\rho^{_B} < 0$, 
$\rho_0 + (1-\Psi_0)^{3/2} \rho^{_B}_0 \geq 0$
offers novel features, since $\rm{P}_2$ is now a saddle point, and 
the dynamics is ruled by the attractor $\rm{P}_3$ instead 
(Figure \ref{plots_rad} bottom left). 
The branes do still start close by ($\rm{P}_1$)
or very far apart ($\rm{P}_2$) but they all eventually move to a finite 
separation ($\rm{P}_3$), determined  which does not bring about the limit of
general relativity. 
In the case 2) with $\rho^{_B} <0$, 
$\rho_0 + (1-\Psi_0)^{3/2} \rho^{_B}_0 \leq 0$ 
there is no attractive fixed point (Figure \ref{plots_rad} bottom right).
Close branes initially separate ($\rm{P}_1$), but then meet again at a 
collision. 

Let us point out that taking the B-brane matter to be zero, i.e.,
considering a scalar-tensor theory with a single usual matter source,
corresponds to the case 1) of radiation domination and 
case 1) of matter domination, yielding qualitatively identical 
phase portraits in the respective cases.
These can be compared with the investigations in general scalar-tensor
cosmology \cite{dn,sa1, skw2} carried out in the Einstein frame and 
using an interpretation of the scalar field in the ``master equation'' 
as a coordinate of a 
fictitious particle with velocity dependent mass. Our results agree -- 
in the radiation domination regime the scalar field slows down and stops
due to friction, while in the matter dominated regime it experiences 
attraction towards general relativity. However, the phase space method 
delivers an easy to understand global picture 
without making any approximating assumptions and 
invoking the unphysical Einstein frame.


\section{Observational constraints}


\subsection{The Eddington parameters and Solar system experiments}

The present state of the cosmological scalar field
can be detected in the Solar system experiments  \cite{bp}. 
Assuming that the B-brane matter has no significant influence on 
these experiments and taking into account Eq. (\ref{ksw}) 
for $\omega (\Psi)$, the observed
values of the gravitational constant $G$ and the
Eddington parameters $\gamma, \beta$ 
read 
\be \label{G(t)}
8 \pi G \equiv \frac{\kappa^2}{\Psi_0} \frac{(4+2\omega_0)}{(3+2\omega_0)}
  = \kappa^2 \frac{4 - \Psi_0}{3\Psi_0} \,,
\ee
\be \label{gam} 
\gamma - 1 \equiv - \frac{1}{\omega_0 +2} = 
- \frac{2(1 - \Psi_0)}{4 - \Psi_0} \,,
\ee
\be \label{bet}
\beta - 1 \equiv  \frac{\kappa^2}{8 \pi G} 
\frac{\omega^\prime(\Psi_0)}{(4+2\omega_0)(3+2\omega_0)^2} =
\frac{2(1-\Psi_0) \Psi_0}{(4-\Psi_0)^2}\,.
\ee
Here $\Psi_0$ denotes
the  cosmological value of the scalar field asymptotically far from Sun
and $\omega_0 \equiv \omega(\Psi_0)$.
In the case $\Psi_0 = 1$ we get the general 
relativistic values $\beta = 1 = \gamma$ 
and $8 \pi G = \kappa^2$. But  in general
$\Psi_0 \not= 1$ and observational limits for $1 -\Psi_0 $ 
can be determined.

The Lunar Laser Ranging experiment (LLR) \cite{llre} gives the constraint 
$4 \beta - \gamma - 3 = (-0.7 \pm 1.0 ) \times 10^{-3}$
and the Shapiro time delay measured by tracking of the Cassini 
spacecraft \cite{cassini} gives 
$ \, \gamma - 1  = (2.1 \pm 2.3) \times 10^{-5} \, $.
In our model $1 -\Psi_0 >0$ and this  implies 
$4 \beta - \gamma - 3 > 0$,   $\gamma - 1<0$. 
It follows that in this case the observational bounds are in fact
$4 \beta - \gamma - 3 < 0.3 \times 10^{-3}$ and 
$1 - \gamma < 2 \times 10^{-6}$ (cf. \cite{palma}). They yield 
observational limits for the asymptotic value of the scalar field 
and the modification of the gravitational constant correspondingly 
\be
  1 - \Psi_0 <  10^{-4}   \quad {\rm  (LLR)}, \qquad
  1 - \Psi_0 < 3 \times 10^{-6} \quad {\rm  (Cassini)} \,;
\ee 
\be
  \kappa^2 / 8 \pi G < 1 - \frac{4}{3}\times 10^{-4} 
\quad {\rm  (LLR)}, \qquad
  \kappa^2 / 8 \pi G < 1 - 4 \times 10^{-6} \quad {\rm  (Cassini)}  \,. 
\ee 
The constraint on the present variation of $\Psi(t)$ is even
more stringent. From the general expression \cite{bp}
\be
\frac{{\dot G}}{G} \equiv - {\dot \Psi}_0 \frac{3 + 2 \omega_0}{4 + 2 \omega_0}
\left( G + \frac{2 \omega^\prime (\Psi_0)}{(3 + 2 \omega_0)^2} \right)
= - \frac{4{\dot \Psi}_0}{\Psi_0 (4 - \Psi_0)}
\ee
and observational data \cite{bp} $|{\dot G} / G| \leq 10^{-11} {\rm yr}^{-1}$
we get 
\be
|{\dot \Psi}_0 / \Psi_0| \leq 10^{-11} {\rm yr}^{-1} \,.
\ee 
We can conclude that the asymptotic background for Solar system
experiments is very near to general relativity $\Psi = 1$, ${\dot \Psi} =0$.


\subsection{Nucleosynthesis  constraints}

Primordial big-bang nucleosynthesis (BBN) provides a probe 
of a very early Universe 
and gives through primordial (relict) light element abundances 
(cosmologically relevant are D, $^3$He, $^4$He, $^7$Li) 
additional constraints for the parameters of the cosmological model 
under discussion. 
It can be used to test the non-standard cosmology as well 
as the non-standard particle physics.

In the framework of  general relativity and 
the standard model of particle physics
the relict abundances depend on  nuclear reaction rates $R_{i}$ 
and only one cosmological parameter, the ratio of baryon number density
$n_b$ to photon number density $n_\gamma$, 
$\eta_{10} \equiv 10^{10} \,\left(\frac{n_{b}}{n_{\gamma}}\right)$.
Here we will not discuss here the reaction rate uncertainties $\Delta R_{i}$, 
for a recent review, see  \cite{cyburt}.
Modifications of the standard cosmological model and the standard
model of particle physics can be given in terms of two 
additional parameters:
1) the lepton asymmetry parameter (asymmetry between 
neutrinos and antinetrinos or neutrino degeneracy)
  $\xi_e$ {\cite{steigman}}; 2) the speed-up factor
$\xi (t) = H(t) / H_G (t)$ (in the framework of our model
we considered it  in Subsect. 4.2 
for the radiation dominated Universe).

Kneller and Steigman \cite{kneller} have presented 
the primordial abundances \{$y_{_D}$, $Y_{_P}$, $y_{Li}$\} 
of \{D,\, $^4$He,\, $^7$Li\} as linear functions of parameters
\{$\eta_{10}$,  $\xi$, $\xi_{e}$\} where $\xi = \xi(t_{nucleosynth}$).
The accuracy of these linear fits is approximately a few 
percent (lower than the current errors in observationally inferred value) 
over realistic ranges of  parameters
  \cite{steigman, kneller}
  $\, 4 \leq \eta_{10} \leq 8 \,$, 
$ \, 0.85 \leq \xi \leq 1.15 \,$,
   $\, -0.1 \leq \xi_{e} \leq 0.1 \,$.
The baryon to photon ratio $\eta_{10}$ is determined at high precision 
from   independent (non-BBN)  measurements of CMB made by WMAP. 
In the standard  $\Lambda$CDM theoretical model its value is
$\eta_{10} = 
6.14 \pm 0.25$ \ 
and it is approximately the same also in the case of nonstandard
cosmological models \cite{kneller}.
We can take into account this (non-BBN) value and observed abundances of 
three elements, but then we cannot adjust the remaining two parameters 
\{$\xi$, $\xi_e$\}
so that all three fits are satisfied. Since the lithium abundance
derived from observations has possible problems, 
Kneller and Steigman \cite{kneller}
derive possible values of the parameters only from 
D and $^4$He abundances in two cases:
\begin{itemize} 
\item if $\xi_e =0$, then $\eta_{10} = 5.88^{+0.64}_{-0.50}$ and $\xi =
0.969^{+0.008}_{-0.009}\,$;

\item if $\eta_{10} = 6.14 \,$, 
then $\xi_e \approx 0.037$ and  $\xi \approx 1.0203 \, $. 
\end{itemize}

Now let us proceed with our specific scalar-tensor theory 
with coupling function (\ref{ksw}).
The nucleosynthesis takes place at the beginning of the radiation 
dominated era and then the influence of the 
cosmological constant is negligible, 
so the relevant speed-up factor is given by Eq. (\ref{speedC}). 
The range of realistic values \cite{kneller} 
$ \, 0.85 \leq \xi \leq 1.15 \,$
translates for $C / \rho_0$ as
\be
  -0.28 < \frac{C}{\rho_0} < 0.32 \,
\ee
(small corrections from $\, \kappa^2 /8 \pi G \,$ ignored)
and above-given observational constraints  \cite{kneller} on $\xi$ yield 
\begin{itemize}
\item for $\xi \approx 0.969\,$ we get $C / \rho_0 \approx -0.063\, $;

\item for $\xi \approx 1.0203$ we get  $C / \rho_0 \approx 0.041\, $. 
\end{itemize}

The initial value of the dark energy density as a 
fraction of the background radiative matter energy density
$C/ \rho_0$ 
is sometimes parametrized by the number of 
extra relativistic degrees of freedom, conventionally 
represented as  additional neutrino 
flavors, $\Delta N_{\nu} \equiv (N_{\nu} - 3)  \, $.
In this case the speed-up factor reads
  \cite{kneller}
\be
\xi^2 = \frac{\kappa^2}{8\pi G} \ \left(1 + \frac{7}{43} \, \Delta 
N_{\nu} \right) 
\qquad \Longrightarrow \qquad
   \frac{C}{\rho_0}=   \frac{7}{43} \Delta N_{\nu} \, \,. 
\ee
Note that in our model the dark radiation is of geometric nature
and its parametrization in terms of additional neutrino
flavors is absolutely formal.

Using updated predictions for light element abundances, BBN and
CMB limit 
Cyburt et al. \cite{cfo, cfos} infer a combined limit 
$ - 0.33 \leq \Delta N_{\nu} \leq 0.85$, meaning 
\be \label{OC}
-0.054 \leq  \frac{C}{\rho_0} \leq 0.138 \,.
\ee
This result can be compared with limits given by Ichiki et al.
\cite{iykom} for the ratio of dark and visible radiation
in the framework of an one-brane model in an approximation
$\rho^2 \approx 0$, since in this case $H^2$ and $\xi^2$ coincide
with the corresponding expressions in our two-brane model.
Ichiki et al. \cite{iykom}  derived from BBN and CMB 
$\,  -0.41 \leq  \frac{C}{\rho_0} \leq 0.105 \, $ 
(using
$\,  4.73 \leq \eta_{10} \leq 5.56 \,$).
We see that limits for the positive dark energy are
approximately the same, but  the negative dark energy density
is much more constrained by Eq. (\ref{OC}).

Finally let us note that the observational limits (\ref{OC}) imply constraints 
on the parameters of the cosmological model in the radiation dominated era, as
\be
\frac{C}{\rho_0} = \frac{\rho_0 + \rho_0^{_B} (1-\Psi_0)^2}{\rho_0 \Psi_{\infty}} - 1
\ee
from Eq. (\ref{C_and_constants}).
For example, we can determine immediately that the case 2) with 
$\rho^{_B} <0, \rho_0 + (1-\Psi_0)^2 \rho^{_B}_0 \leq 0$, discussed in
Subsec. \ref{rad_dom_dyn_sys}, is by far ruled out by the observations.


\section{Concluding remarks}


In this paper we have investigated a scalar-tensor cosmology 
arising as an effective 4-dimensional description
of the 5-dimensional Randall-Sundrum two-brane cosmological scenario \cite{rs1} 
in the Kanno-Soda low energy gradient expansion approximation \cite{ks1}. 
The model is characterized by a scalar field $\Psi$ (radion)
measuring the proper distance of the branes and endowed with 
a specific coupling function (\ref{ksw})
implied by the construction of the model, 
a specific potential (\ref{VPsi}) generated by cosmological
constants on the branes,
and barotropic fluid matter sources on both branes.
In the limit when the (hidden) B-brane is empty, the model reduces
to a scalar tensor cosmology with cosmological constant and matter,
the limit of general relativity is obtained when $\Psi=1, \dot{\Psi}=0$.
We have assumed FLRW metric on both branes.

Due to the specific form of the coupling function the Jordan frame 
equations of motion combine to give a scale factor dynamical equation
decoupled from the scalar field and B-brane matter, 
leading to a Friedmann equation with only an additional dark radiation
term reminding the extra dimension.
This has enabled us to find exact analytic solutions 
for the flat ($k=0$) Universe scale factor with various kinds 
of matter: cosmological constant, radiation, dust, and cosmological 
constant plus radiation. We have also found analytic solutions for
the scalar field with radiation, and with cosmological constant plus  radiation
(latter case to be expressed in terms of elliptic functions).
Over the years Barrow et al. have developed methods to obtain 
exact analytic solutions for scalar tensor cosmologies with general 
coupling functions \cite{bp, Barrowetal}. 
Forgetting the B-brane component, our model belongs to  
classes 1 and 3 studied in ref. \cite{bp}.
However, in the general case the solutions are not explicitly available in the
cosmic time and one resorts to approximation techniques for early and late 
time behavior. Therefore, despite being a specific one, our model has
an advantage of being directly solvable, 
thus providing an interesting example
of a scalar-tensor cosmology throughout the entire radiation, dust 
and cosmological constant dominated eras.

Although our model is governed by the Friedmann equation (with an extra 
dark radiation term) and hence in broad accordance with standard cosmology,
we need to study the behavior of the scalar field, 
in order to be assured that the theory satisfies all observational constraints 
and converges towards general relativity.
A clever reparametrization of time \cite{sa1} allows to extract an 
independent evolution equation for the scalar field also when besides
matter on the A-brane, there is similar matter ($\Gamma^{^B} = \Gamma$) 
on the B-brane. We examine the solutions of this ``master equation''
by applying
the methods of dynamical systems. It seems 
determining the fixed points and drawing
the phase portraits is a more straightforward approach than the usual 
procedure of changing into the unphysical Einstein frame and relying on 
an analogue with a fictitious particle with velocity dependent mass.
Still, in the absence of B-brane matter, the results are 
in accord with previous general studies \cite{dn,skw2,sa1}:
in a flat Universe with positive cosmological constant or dust 
the system relaxes to the limit of general relativity, while in the radiation
case friction makes the scalar field to stop at an arbitrary position.

Inclusion of the B-brane matter (possibly having negative energy density)
complicates the story, as was also previously 
noted from the numerical studies in the 
moduli space approximation approach \cite{bbdr}.
In summary, we find that general relativity, 
which in the branes' picture corresponds to
infinitely far away branes, is the sole destiny for 
cosmological constants $\sigma + \sigma^{_B}\geq0$, or dust 
$\rho_0 \geq 0$, $\rho_0^{_B}\geq 0$. 
The solutions may approach general relativity or go the other way, 
i.e., to brane collision, depending on the initial conditions, for 
cosmological constants $\sigma + \sigma^{_B}\leq0$, or radiation 
$\rho_0 + (1-\Psi_0)^2 \rho^{_B}_0 \leq 0$.
Brane collision is the only future for dust with 
$\rho_0 + (1-\Psi_0)^{3/2} \rho^{_B}_0 \leq 0$.
Radiation with $\rho_0 + (1-\Psi_0)^2 \rho^{_B}_0 \geq 0$
has the radion stabilizing at arbitrary position due to friction. 
There are also fixed points in between the two extremes, namely 
a saddle for cosmological constants $\sigma>0$, $\sigma + \sigma^{_B}\leq0$,
and an attractor for dust 
$\rho^{_B}_0<0$, $\rho_0 + (1-\Psi_0)^{3/2} \rho^{_B}_0 \geq 0$.
The latter is especially interesting, since we see a natural stabilization
of the radion for which one usually invokes an additional scalar field 
in the bulk \cite{gw}. 

The solar system tests restrict the present value of the radion to
$1 - \Psi <  10^{-4}$,
and allow it to be only very slowly varying, 
$|{\dot \Psi} / \Psi| \leq 10^{-11} {\rm yr}^{-1}$. 
The observations of D and He abundances together with CMB
set limits on the fraction of dark radiation to normal radiation,
$-0.054 \leq  \frac{C}{\rho_0} \leq 0.138$. 
The results of these experiments can be accommodated
in the model if its parameters are correspondingly constrained, 
which is not too hard, given that 
for large range of parameters general relativity is an attractor. 
Due to the specific coupling function (\ref{ksw}) the first integral 
(\ref{H^2}) for the two-brane cosmology coincides with the Friedmann 
equation for the  one-brane cosmology \cite{langl} in the approximation 
where the matter density $\rho$ on the brane is small and quadratic term 
$\rho^2$ can be neglected. 
This means that observational constraints for our model can be analyzed 
along the same patterns as in the case of the one-brane cosmology 
\cite{iykom, bgsw}.
Also note that the specific coupling (\ref{ksw}) implies constant speed-up 
factor (\ref{speed}) in the radiation dominated era and excludes a variable 
speed-up factor mechanism recently proposed by Larena et al. \cite{larena} 
for solving the Li abundance problem.



\begin{thebibliography}{99}

\bibitem{rs1}
L. Randall and  R. Sundrum,  
Phys. Rev. Lett. 83: 3370 (1999),  
[hep-ph/9905221].

\bibitem{rs2}
L. Randall and R. Sundrum, 
Phys. Rev. Lett. 83: 4690 (1999), 
[hep-th/9906064].

\bibitem{lukas}
A. Lukas, B.A. Ovrut, K.S. Stelle, and  D. Waldram, 
Phys. Rev. D 59: 086001 (1999), 
[hep-th/9803235].

\bibitem{ekpyrotic}
J. Khoury, B. Ovrut, P.J. Steinhardt, and  N. Turok, 
Phys. Rev. D 64: 123522 (2001), 
[hep-th/0103239].

\bibitem{ksw}
S. Kanno, J.Soda, and D. Wands, 
JCAP 0508: 002 (2005),  
[hep-th/0506167].

\bibitem{ks2}
S. Kanno and J. Soda, 
Phys. Rev.  D 66: 043526 (2002),
[hep-th/0205188].

\bibitem{ks1}
S. Kanno and J. Soda, 
Phys. Rev. D 66: 083506 (2002), 
[hep-th/0207029].

\bibitem{bbdr}
Ph. Brax, C. van de Bruck, A.-C. Davis, and C.S. Rhodes, 
Phys. Rev. D 67: 023512 (2003), 
[hep-th/0209158].

\bibitem{kanno}
S. Kanno,
Phys. Rev. D 72: 024009 (2005), 
[hep-th/0504087].

\bibitem{dn}
T. Damour and K. Nordtvedt,
Phys. Rev.  D 48: 3436 (1993).

\bibitem{sa1}
A. Serna, J.M. Alimi, and A. Navarro, 
Class. Quant. Grav.  19: 857 (2002),
[gr-qc/0201049].

\bibitem{skw2}
D.I. Santiago, D. Kalligas, and R.W. Wagoner,
Phys. Rev.  D 58: 124005 (1998),
[gr-qc/9805044].

\bibitem{bp}
J.D. Barrow and P. Parsons, 
Phys. Rev.  D 55: 1906 (1997),
[gr-qc/9607072].

\bibitem{skw1}
D.I. Santiago, D. Kalligas, and R.W. Wagoner,
Phys. Rev.  D 56: 7627 (1997),
[gr-qc/9706017].

\bibitem{dp} 
T. Damour and B. Pichon,
Phys. Rev.  D 59: 123502 (1999),
[astro-ph/9807176].

\bibitem{cbs}
T. Clifton, J.D. Barrow, and R.J. Scherrer,
Phys. Rev. D 71: 123526 (2005),
[astro-ph/0504418].

\bibitem{coc}
A. Coc, K.A. Olive, J.-P. Uzan, and E. Vangioni,
Phys. Rev. D 73: 083525 (2006),
[astro-ph/0601299].


\bibitem{palma}
G.A. Palma, 
Phys. Rev. D 73: 044010 (2006),
[hep-th/0511170].

\bibitem{Faraoni_book}
V. Faraoni, 
``Cosmology in scalar tensor gravity'', 
Dordrecht, Netherlands: Kluwer Academic Publishers (2004).

\bibitem{kss}
S. Kanno, M. Sasaki, and J. Soda, 
Prog. Theor. Phys.  109: 357 (2003), 
[hep-th/0210250].

\bibitem{iii}
P. Kuusk and M. Saal, 
Gen. Rel. Grav.   36: 1001 (2004), 
[gr-qc/0309084].

\bibitem{RS_cosmo_all} 

C. Csaki, M. Graesser, C.F. Kolda, and J. Terning,
Phys. Lett. B 462: 34 (1999),
[hep-ph/9906513].

J. M. Cline, C. Grojean, and G. Servant,
Phys. Rev. Lett.  83: 4245 (1999),
[hep-ph/9906523].

P. Binetruy, C. Deffayet, U. Ellwanger, and D. Langlois,
Phys. Lett. B 477: 285 (2000),
[hep-th/9910219].

C. Csaki, M. Graesser, L. Randall, and J. Terning,
Phys. Rev. D 62: 045015 (2000),
[hep-ph/9911406].

P. Binetruy, C. Deffayet, and D. Langlois,
Nucl. Phys. B 615:  219, (2001),
[hep-th/0101234].

K. Koyama,
Phys. Rev. D 66: 084003 (2002),
[gr-qc/0204047].

S. Kobayashi and K. Koyama,
JHEP 0212: 056 (2002),
[hep-th/0210029].

I. Brevik, K. Borkje, and J.P. Morten,
Gen. Rel. Grav.  36: 2021 (2004),
[gr-qc/0310103].

J. Martin, G.N. Felder, A.V. Frolov, M. Peloso, and L. Kofman,
Phys. Rev. D 69: 084017 (2004),
[hep-th/0309001].

J. Martin, G.N. Felder, A.V. Frolov, L. Kofman, and M. Peloso,
Comput.  Phys. Commun.  171: 69 (2005),
[hep-ph/0404141].

A. Wang, R.G. Cai, and N.O.Santos,
[astro-ph/0607371].

\bibitem{palma_davis_MSA}
G.A. Palma and A.C. Davis,
Phys. Rev. D 70: 064021 (2004),
[hep-th/0406091].

\bibitem{webster_davis_MSA}
S.L. Webster and A.C. Davis,
[hep-th/0410042].

\bibitem{garriga}
J. Garriga and T. Tanaka,
Phys. Rev. Lett. 84: 2778 (2000),
[hep-th/9911055].

\bibitem{langl}
P. Binetruy, C. Deffayet, U. Ellwanger, and D. Langlois,
Phys. Lett. B 477: 285 (2000),
[hep-th/9910219].

\bibitem{dark_acceleration}

E. Kiritsis, G. Kofinas, N. Tetradis, T.N. Tomaras, and V. Zarikas,
JHEP 0302: 035 (2003),
[hep-th/0207060].

N. Tetradis,
Phys. Lett. B 569: 1 (2003),
[hep-th/0211200].

K. I. Umezu, K. Ichiki, T. Kajino, G.J. Mathews, R. Nakamura, and M.Yahiro,
Phys. Rev. D 73: 063527 (2006),
[astro-ph/0507227].

P.S. Apostolopoulos and N. Tetradis,
Phys. Lett. B 633: 409 (2006),
[hep-th/0509182].

\bibitem{close_branes}

T. Shiromizu, K. Koyama, and K. Takahashi,
Phys. Rev. D 67: 104011, (2003),
[hep-th/0212331].

C. de Rham and S. Webster,
Phys. Rev. D 71: 124025 (2005),
[hep-th/0504128].

C. de Rham and S. Webster,
Phys.Rev. D 72: 064013 (2005),
[hep-th/0506152].

C. de Rham, S. Fujii, T. Shiromizu, and H. Yoshino,
Phys. Rev. D 72: 123522 (2005),
[hep-th/0509194].

\bibitem{llre}
J. G. Williams, X. X. Newhall, and J. O. Dickey,
Phys. Rev.  D 53:  6730 (1996).

\bibitem{cassini}
B. Bertotti, L. Iess, and P. Tortora,
Nature 425: 374 (2003).

\bibitem{cyburt}
R. Cyburt,
Phys. Rev.  D 70: 023505  (2005),
[astro-ph/0401091].

\bibitem{steigman}
G. Steigman,
Int. J. Mod. Phys. E 15: 1 (2006),
[astro-ph/0511534].

\bibitem{kneller}
J.P. Kneller and G. Steigman,
New J. Phys. 6: 117 (2004),
[astro-ph/0406320].

\bibitem{cfo}
R.H. Cyburt, B.D. Fields, and K.A. Olive,
Phys. Lett. B 567: 227 (2003),
[astro-ph/0302431].

\bibitem{cfos}
R.H. Cyburt, B.D. Fields, K.A. Olive, and E. Skillman,
Astropart. Phys. 23: 313 (2005),
[astro-ph/0408033].

\bibitem{iykom}
K. Ichiki, M. Yahiro, T. Kajino, M. Orito, and G.J. Mathews,
Phys. Rev.  D 66: 043521 (2002),
[astro-ph/0203272].

\bibitem{bgsw}
J.D. Bratt, A.C. Gault, R.J. Scherrer, and T.P. Walker,
Phys.Lett. B 546: 19 (2002),
[astro-ph/0208133].

\bibitem{Barrowetal}
J.D. Barrow, K. Maeda,
Nucl. Phys. B 341: 294 (1990).

J.D. Barrow, 
Phys. Rev. D 47: 5329 (1993).

J.D. Barrow, J. Mimoso, 
Phys. Rev. D 50: 3746 (1994).

\bibitem{gw}
W.D. Goldberger, M.B. Wise,
Phys. Rev. Lett. 83: 4922 (1999),
[hep-ph/9907447].

\bibitem{larena}
J. Larena, J.-M. Alimi, and A. Serna,
(submit. to Phys. Rev. D),
[astro-ph/0511693].

\end{thebibliography}
\end{document}